\documentclass[12pt]{article}

\usepackage{amsmath,amsfonts,epsfig,color,latexsym}

\newcommand{\be}{\begin{equation}}
\newcommand{\ee}{\end{equation}}
\newcommand{\bea}{\begin{eqnarray}}
\newcommand{\eea}{\end{eqnarray}}

\renewcommand{\theequation}{\thesection.\arabic{equation}}

\let\newsection=\section
\renewcommand{\section}{\setcounter{equation}{0}\newsection}

\begin{document}

\begin{flushright}
hep-th/0501013\\
BROWN-HET-1427
\end{flushright}
\vskip.5in

\begin{center}

{\LARGE\bf Towards S matrices on flat space and pp waves from SYM}
\vskip 1in
\centerline{\Large Antal Jevicki and Horatiu Nastase}
\vskip .5in

\end{center}
\centerline{\large Brown University}
\centerline{\large Providence, RI, 02912, USA}

\vskip 1in

\begin{abstract}
{\large
We analyze the possibility of extracting S matrices on pp waves and flat 
space from SYM correlators. For pp waves, there is a subtlety in defining 
S matrices, but we can certainly obtain observables. Only extremal correlators
survive the pp wave limit. A first quantized string approach is 
inconclusive, producing in the simplest form 
results that vanish in the pp wave limit. 
We define a procedure to get S matrices from SYM correlators, both for flat 
space and for pp waves, generalizing a procedure due to Giddings. We analyze 
nonrenormalized correlators: 2 and 3 -point functions and extremal 
correlators. For the extremal 3-point function, the SYM and AdS results 
for the S matrix match for the angular dependence, but the energy dependence
doesn't.

}

\end{abstract}

\newpage

\section{Introduction}

One can understand holography rather straightforwardly in the usual AdS-CFT 
correspondence \cite{malda}. The boundary of global $AdS_5\times S^5$ 
space is $S^3\times 
R_t$. Supergravity correlators in the bulk, with sources living on the 
boundary, are the same as SYM correlators on the boundary \cite{witten,gkp}. 
SYM operators 
on the $R^4$ plane correspond by conformal invariance to states on $S^3\times
R_t$, and are mapped by AdS-CFT to normalizable modes in 
global $AdS_5\times S^5$. 

The size of $AdS_5\times S^5$ in string units 
is given by the 't Hooft coupling, $R/\sqrt{\alpha'}=(g_{YM}^2N)^{1/4}$, 
and thus can be varied. Increasing the size of the space will lower string 
$\alpha '$ corrections, but given that the AdS observables are defined by  
putting sources on the boundary, it would seem that we will always feel 
the curvature of space (expressed by the SO(d,2) invariance of $AdS_{d+1}$,
mapped to conformal invariance on the boundary). Yet it should also 
(intuitively) be possible to obtain a flat space limit, in which in 
particular one should be able to define S matrices. There were several 
proposals for how to do this \cite{polc,suss,bgl,gidd,gid}, 
but all suffered from some problems, 
basically stemming from the difficulty to eliminate the contribution of 
the boundary in AdS observables, and keep only contribution from a (quasi)flat
region of space in the middle of the bulk. 

On the other hand, it was found in \cite{bmn} that one can take a particular 
limit on the $AdS_5\times S^5$ space, the Penrose limit, that focuses 
in on a null geodesic sitting in the middle of $AdS_5$ and spinning around on 
the $S^5$, and understand it from SYM. The point was that we need to take a 
subsector of SYM, involving operators of large R charge. The issue of 
holography though is very tricky, since the original AdS-CFT has no reason 
to apply anymore, as we are focusing in on a region far away from the AdS 
boundary. The issue generated some confusion \cite{dgr,kp,lor}, 
but in \cite{bn} it was 
pointed out that for the pp wave correspondence holography 
works in a different way: One has a map between SYM states on $S^3\times R_t$,
dimensionally reduced on $S^3$, and the string worldsheet, identifying the 
SYM time t with the worldsheet time $\tau$ of the string, in first quantization
or second quantization (string field theory). Many calculations followed 
exploring this correspondence (\cite{cfhmmps,cfhm,kpss,bkpss,sv,svtwo,psvvv},
... for a more complete list see e.g. \cite{ds}), 
and getting agreement between SYM calculations 
and string field theory calculations in the pp wave. Thus holography is 
in this case of a radically different type, linking SYM theory to 
worldsheet string theory. A different approach was also tried 
\cite{dsy,dy,dytwo}, in 
which holography on the pp wave acts through instantonic (complex) paths 
that do reach the AdS boundary in the pp wave limit, but that also has 
its problems.

On the other hand, now that we are actually in the middle of AdS, it is 
conceivable that we could more easily define S matrices, a fact that we 
saw was hard before (for taking directly the flat space limit). In fact, 
as was already noticed in \cite{bn}, a simple order of magnitude 
calculation shows agreement between an amplitude obtained from SYM 
(${\cal A}\sim J^{3/2}/N \delta_{\sum J_i, \sum J'_i}$)
and amplitudes (vertices) of S matrix type, i.e. obtained from 
the pp wave supergravity  acting on normalizable wavefunctions (${\cal A}
\sim g_s p^{3/2}\delta (\sum p_i-\sum p'_i)$). It therefore seems likely 
that we can define S matrices on pp waves and derive them from SYM.
The question of defining S matrices on pp waves was addressed in \cite{bs} and 
it was found that there seems to be a consistent definition at tree level.

There is one more problem with this fact though
\footnote{We thank Juan Maldacena for pointing this out to us. See also 
the Note added in \cite{bs}}. Namely, there is a subtlety in the definition of 
S matrices for massless external states. On the pp wave, string states 
have
\be
-2p^-= \mu \sum_i N_{0i}+\mu \sum _i \sum _{n_i}\sqrt{1+\frac{n_i^2}{
(\mu\alpha ' p^+)^2}}= \sum_i \frac{p_i^2}{p^+}+\frac{M^2}{p^+}
\label{stringstates}
\ee
where we have rewritten the first term in terms of discrete momenta 
$p_i$ that will become continous in the flat space limit of the pp wave 
($\mu\rightarrow 0$).
To construct the S matrix we first construct wavepackets 
\be
\phi = \sum _{\vec{n}}\int \frac{dp^-}{2\pi} \int \frac{dp^+}{\sqrt{2p^+}}
\phi_{\vec{n}}(p^+, p^-) \phi_{\vec{n}, p^+, p^-}(x^+, x^-, x^i)
\ee
where $\phi_{\vec{n}, p^+, p^-}(x^+, x^-, x^i)$ are the eigenmodes of the wave 
operator, or equivalently 
\be
|\phi> = \sum _{\vec{n}}\int \frac{dp^-}{2\pi} \int \frac{dp^+}{\sqrt{2p^+}}
\phi_{\vec{n}}(p^+, p^-)|\vec{n}, p^+, p^->
\ee

The boundary of the pp wave space is a one dimensional null line, parametrized
by $x^+$ \cite{bn}, and so asymptotic states can only be defined in one 
direction ($x^-$), parametrized by $p^+$. Indeed, in the transverse directions
($x^i$), there is a harmonic oscillator potential, thus states are just 
parametrized by oscillator numbers, but are not asymptotic, and then the 
energy $p^-$ is derived on-shell from them. Thus \cite{bs} define S matrices
as scatterings in a two dimensional effective theory ($p^+,p^-$), with 
extra discrete indices (oscillators):
\be
{\cal L}_2= \sum_{\vec{n}}\phi_{\vec{n}}^*(p^+, p^-)(p^--{\cal E}(\vec{n},
p^+))\phi_{\vec{n}}(p^+, p^-)+{\cal L}_{int}
\ee
Note though that now the dispersion relation (\ref{stringstates}) at $M=0$ 
has no $p^+$ dependence (in the form after the first equal sign), 
i.e. $\partial p^-/\partial p^+=0$
so no group velocity, thus one cannot kick the waves \cite{bs}. Another way 
to spot the problem at M=0 is to try to write down cross sections using the 
wavepackets. Following what happens in flat space when we use wavepackets to 
turn S matrices to cross-sections, we see the problem: S matrices
have overall momentum delta functions $\delta ^d(\sum p_i)$, but external 
particles are on shell, thus integrations are over $d^{d-1}k_j$. Usually,
the last integration in $d\sigma $ is of the type
\be
\int dp_1 \frac{p_1^{d-2}d\Omega}{2E_{1, \vec{p}}}\delta (E_{1, \vec{p}}-...)
...
\ee
where the last delta function remains from the S matrices. If E is a function 
of $\vec{p}$, then the delta function is cancelled, but if not, like 
for the massless S matrices in our case, the delta function  
remains in the final result for $d\sigma/d \Omega$, making it infinite. 

Our point of view is that first of all, even if it is multiplied by an 
infinity, the 
massless S matrix thus defined is still an observable, that one can try to
get from SYM. Second, we can always mimic what happens in the flat space 
limit of the pp wave, as we did in the second equality in (\ref{stringstates}).
That is, write down wavepackets in the discrete momenta $p_i=\sqrt{\mu p^+ 
N_{0i}}$ that localize particles in the $x^i$ directions. As a perturbation 
around flat space, this should be possible, although it is not clear 
if it is possible in general (this would amount to having asymptotic states).
If this procedure is valid, it would generate the effective $p^+$ dependence 
for the massless case in the second line of  (\ref{stringstates}).

In this paper, we will try to get the S matrices thus defined from SYM 
correlators, using a generalization of 
the procedure put forward by Giddings \cite{gid} in 
the massless AdS case. To our knowledge, there was no direct test of 
the procedure available, so we will also try to do the simplest case.

We will first explain and expand on ideas from \cite{bn}, in section 2. 
Specifically, we will show that as far as extracting pp wave S matrices 
from SYM, only extremal correlators survive, all other correlators becoming 
subleading in the limit (cannot be expressed in terms of pp objects alone). 
This is similar in spirit to the fact that in the limit, only certain operators
(BMN) survive from SYM. 

In section 3, we will show that the first quantized string 
calculation in \cite{mp}, while correct, 
gives a result that is vanishing in the Penrose limit, and again only 
extremal correlators will survive. In \cite{mp}, the correlators $<I^2|{\cal O}
^{I_3}|I_1>$ were analyzed on both AdS and SYM sides, and interpreted 
as a pp wave string amplitude for a state $|I_1>$ (with $k_1=\tilde{k}_1
+J$ in SYM) to go into a state $|I_2>$ (with $k_2=\tilde{k}_2+J$),
by propagating in the perturbed metric produced by $O^{I_3}$ (with 
$k_3=\tilde{k}_3$), but this object vanishes in the Penrose limit. 
We say how one could obtain a nonzero result and show that this forces 
us towards the string field theory calculation, where $p^+$ changes, i.e. 
$J_3=J_1-J_2$ of the same order as $J_2$. In that case, extremality 
of SYM correlators is forced upon us, as conservation of pp wave momentum 
$p^+$. 

In section 4 we will explain the Giddings procedure for extracting S matrices
from SYM correlators, and generalize it to the case of nonzero AdS mass and 
then to pp waves. Basically, one needs to turn boundary to bulk propagators 
in global AdS into normalizable wavefunctions. We show that AdS wavefunctions 
have the correct flat space and pp wave limits, and also that pp wave 
wavefunctions go over to flat space wavefunctions. 

In section 5 we set out to test the procedure using the correlators known to 
be non-renormalized: scalar 2 and 3-point functions and extremal correlators.
The general 2 and 3-point functions are relevant for the flat space limit, 
while the large charge extremal correlators are relevant for the pp wave case. 
We compare in detail the simplest case of extremal 3-point function on 
both sides.

In section 6 we present our conclusions and avenues for future work. 
Appendix A offers an overview of correlators and holography in the 
various coordinates used (Euclidean Poincare, Lorentzian Poincare, 
Lorentzian global, $(t,\vec{u})$ cylinder). Appendix B contains the expansion 
of the delta function in spherical harmonics (for comparison with SYM 
calculations). Appendix C contains identities needed for the limits of 
AdS wavefunctions, and Appendix D contains the general 3-point function 
of scalars.

\section{Extremal correlators and the Penrose limit}

In this section we explain and expand
the points made in \cite{bn} about the importance of 
extremal correlators in the Penrose limit.

In order to define S matrices on the pp wave we need to understand the 
relation between the normalizable modes on the pp wave and states in SYM.

The pp wave metric is
\be
ds^2=2dx^+dx^--\mu^2 r^2 (dx^+)^2 +dx^i dx^i
\ee
where $r^2= x^i x^i$.

The normalized solutions to the wave equation on the pp wave 
\be
(\Box-m^2)
 \phi =(2\partial_+\partial_-+\mu^2r^2\partial_-^2 +\partial_i^2- m^2)\phi=0
\ee
are 
\be
\phi = \phi(x^+)\psi(x^-)\psi_T(x_i)= (e^{ip^-x^+}) (B e^{ip^+x^-})
\prod_i(\frac{(p^+)^{1/4}}{\sqrt{\sqrt{\pi}2^n n!}}e^{-p^+(x_i)^2}
H_{n_i}(\sqrt{p^+}x^i))
\ee
where
\be
-2p^+p^-= 2\mu p^+\sum_{i=1}^8 (n_i+\frac{1}{2})+m^2
\ee
and the normalization constant B depends on the problem. In the gravity 
calculation in SYM, one takes the noncompact  
$B_{NC}=1/\sqrt{p^+}$  delta function (in momentum) 
normalization. But the pp wave limit comes from the (modified) Penrose limit 
of AdS, where $p^+=J/R^2$, as $x^-$ is a circle of radius $R^2$. Therefore, 
when we calculate SYM amplitudes rather than gravity amplitudes, we have 
to use the compact normalization ($\delta_{J_1J_2}$) for states, thus 
\be
B_C=\frac{1}{\sqrt{p^+}R}= \frac{B_{NC}}{R}
\ee
and correspondingly gravity amplitudes (noncompact) with m external 
legs are related to SYM amplitudes (compact) via 
\be
A_{NC}= R^m A_C
\ee
but we also need to remember to change the momentum conservation delta 
function according to 
\be
\delta(p^+_{in}-p^+_{out})=R^2\delta_{J_{in},J_{out}}
\ee

A closed string field $\phi$ in the pp wave background, for definiteness 
a (massless)
graviton mode, will have an interaction (vertex, obtained in this case
from expanding the Einstein action) of the type
\be
g_s \int d^8 r dx^+ dx^- \phi^2 \Box \phi
\ee
that will give a 3-point function that behaves as (keeping only $p^+$ 
factors and dropping the + index)
\be
A_{NC}\sim g_s p^{3/2}\delta (p_3-p_1-p_2)
\ee
It will convert in SYM variables to 
\be
A_C\sim \frac{J^{3/2}}{N}\delta_{J_3, J_1+J_2}
\ee
and note that the $p^{3/2}$ behaviour is the unique behaviour that converts 
into a SYM amplitude, so we chose a representative vertex. Viceversa, 
the $J^{3/2}/N$ behaviour is the unique SYM behaviour that translates into 
a closed string 3-point vertex, without extra powers of R that take us 
away from the Penrose limit (given that we need to have 
$g_s$ and the delta function).

Any SYM 
3 point function that behaves in a different way will not translate into 
a good pp wave amplitude (but rather into an AdS amplitude, away from the 
Penrose limit). 

The class of SYM correlators that have been proven to be 
not renormalized is composed of 3 point functions and so called ``extremal''
scalar correlators, when the number of fields on one operator matches the 
sum of the others, $k_1=k_2+...+k_n$, or more generally $k_1+....+k_n= 
k_{n+1}+...k_{n+m}$, which was conjectured in \cite{dfmmr} to have a 
nonrenormalization theorem. 

Extremal correlators match with AdS calculations only when one takes into 
account a certain analytical continuation. Indeed, the AdS calculation 
for 3 point functions starts with a coefficient for the 3-point vertex
that is zero: $\alpha_3= k_1+k_2-k_3=0$, but in the AdS 3-point function it 
gets multiplied by $1/\alpha_3$ \cite{lttwo}.

When we take the Penrose limit however, the J momenta become continous 
$(p^+)$, and the delicate problem of analytic continuation dissappears.

The 3-point function of scalar operators has been evaluated exactly 
and has a factor of $\sqrt{k_1k_2k_3}/N$, thus if all the $k$'s are of 
order J, the correlator is of order $J^{3/2}/N$ as we claimed that we need,
and there would be no problem. There is also an invariant tensor $<
{\cal C}^{I_1}{\cal C}^{I_2}{\cal C}^{I_3}>$ that would need to be rescaled,
but in the general case it would not bring in new powers of J.  
However, in the Penrose limit we are interested
in operators that have mostly Z insertions (where $Z=\Phi^5+i\Phi^6$
is a complex scalar) and only a few $\Phi^i$ ``impurities'' (i=1,4), and 
then we can have new powers of J. We will also complexify the impurities 
$\Phi= \Phi^1+i\Phi^2, \Phi'=\Phi^3+i\Phi^4$, but we will generically 
write $\Phi$.

There is a simple way to see the $J^{3/2}/N$ behaviour. 
Let's begin with the contraction (overlap
amplitude) of between a 2-trace operator
\be
\frac{Tr(\Phi Z^{J_1})}{\sqrt{N^{J_1+1}}}\frac{Tr(\Phi Z^{J_2})}{\sqrt{N^{J_2
+1}}}(x)
\ee
and the single trace operator (with $J=J_1+J_2$)
\be
\frac{\sum_l \Phi Z^l \Phi Z^{J-l}}{\sqrt{J N^{J+2}}}(z)
\ee
We can see that the non-planar overlap of the two will be of order
\be
\frac{J^2}{N\sqrt{J}}
\ee
where the $N\sqrt{J}$ comes from the normalizations and the $J^2$ comes from 
a choice where to break the two strings to be glued with respect to their
origin. This result was obtained from free SYM overlaps, but this is also what 
happens in AdS-CFT, and here we just have a limit of that calculation.

Notice that we could have taken just as well instead of the double trace
operator two single trace operators at different points,
\be
\frac{Tr(\Phi Z^{J_1})}{\sqrt{N^{J_1+1}}}(x)
\frac{Tr(\Phi Z^{J_2})}{\sqrt{N^{J_2+1}}}(y)
\ee
and the calculation would go through.

We can easily convince ourselves that this calculation also 
goes over if we have 
$a_1$ insertions in the first trace ($J_1$), $a_2$ insertions in the second 
trace ($J_2$) and $a=a_1+a_2$ insertions into the single trace operator, 
the difference in normalizations being compensated by extra factors 
from summations. So in the Penrose limit, these ``extremal'' correlators
all give the correct result.

What happens if we go away from extremality? The simplest example is the 
correlator 
\be
<\frac{Tr(\Phi Z^{J_1}\Phi)}{\sqrt{J_1N^{J_1+2}}}(x)\frac{Tr(\bar{\Phi}
Z^{J_2})}{\sqrt{N^{J_2+1}}}(y) \frac{Tr(\Phi Z^{J_1+J_2})^*}
{\sqrt{N^{J_1+J_2+1}}}(0)>
\ee
where we have dropped the summation in the first operator and wrote the 
fields in the order of contraction (and the neighbouring $\Phi$ and 
$\bar{\Phi}$ are contracted as well). We need to make one contraction between 
operators 1 and 2, thus it is a step away from extremality. 

To remain in the planar limit (and thus have a 1/N interaction) we need to 
keep the $\Phi$ and $\bar{\Phi}$ as they are, and thus the summation 
brings in only a factor of $J$, not $J^2$, and the norm factors are the 
same as before, thus the amplitude is of order $\sqrt{J}/N$, down 1/J
from the previous.

We can easily convince ourselves that all 
3-point non-extremal correlators in 
the Penrose limit will have the same fate, namely they will be subleading,
and as such will not have a gravity interpretation (only away from the
limit, in AdS). Moreover, exactly the same argument can be generalized 
easily to show that only $n+m$-point  extremal correlators 
$k_1+...+k_n=k_{n+1}+...+k_{n+m}$ survive in general. 

In the general extremal correlator case, with $n+m=q$, the closed string 
interaction will be of the type (again, for example, by expanding the 
Einstein action)
\be
g_s^{q-2}\int d^8r dx^+ dx^- \phi^{q-1}\Box \phi
\ee
which will give a gravity amplitude 
\be
{\cal A}_{NC}\sim
g_s^{q-2}\frac{p (p^2)^{q-2}}{p^{q/2}}\delta (\sum_{i=1}^np_i- \sum_{j=1}^m
p_j)
\ee
that will translate into  SYM variables in
\be
{\cal A}_C\sim (\frac{J^{3/2}}{N})^{q-2}\delta_{\sum_{i=1}^nJ_i, \sum_{j=1}^m
J_j}
\ee
and that will be what one gets from the extremal SYM correlators as well.

The flat space limit of the pp wave means $\mu\rightarrow \infty$. In this 
limit, $(p^i)^2= \mu p^+ n_i$ becomes continous and the pp wave supergravity
on-shell relation becomes the flat space one, $2p^+ p^-=\vec{p}^2 + m^2$.

Similarly, the string modes 
\be
-2p^-= \mu \sum_i \sum_{n_i} N_{n_i} \sqrt{1+\frac{n^2}{(\alpha '\mu p^+)^2}}
\ee
become, with the momenta (zero modes) as before
 (just with a change of notation), 
$\vec{p}^2= \sum _i \mu p^+ N_{0,i}$, in the flat space limit
\be
-2p^+p^-= \vec{p}^2 + M^2;\;\;\ M^2= \frac{1}{\alpha '}\sum_i \sum_{n_i}
N_{n_i} n_i
\ee

\section{The first quantized string approach}

In this section, we will show that the calculation in \cite{mp} while
correct, it gives a result that is vanishing in the pp wave limit, and 
the only nonzero result comes from extremal SYM correlators. We will
also see that it is hard to calculate string S matrices using this approach,
but we can say a few general things about it.

In \cite{lmrs}, the 2-point functions of normalized chiral operators 
were given 
\be
< {\cal O} (x_1) {\cal O} (x_2) > =\frac{\delta ^{I_1I_2}}{|x_{12}|^{2k}}
\ee
implying a 3-point function 
\be
< {\cal O} (x_1) {\cal O} (x_2) {\cal O}(x_3) > = \frac{1}{N}
\frac{\sqrt{k_1k_2k_3}<{\cal C}^{I_1}{\cal C}^{I_2} {\cal C}^{I_3} >}
{|x_{12}|^{2\alpha_3}|x_{23}|^{2\alpha_1}|x_{13}|^{2\alpha_2}}
\ee
All of these correlator
calculations correspond, as mentioned, to Euclidean Poincare AdS.
Here $2\alpha_3=k_1+k_2-k_3$, etc.

As we can see, if all the 3 charges $k_i$ are of order J in the large 
J limit, and the SO(6) tensor is of order one, then the 3-point function 
is of order $J^{3/2}/N$. The explicit proof in 
the previous section and in \cite{bn} shows that this 
is the case for BPS extremal operators (with $k_1=k_2+k_3$). 

In \cite{mp}, the case analyzed is $k_1=\tilde{k}_1+J, k_2=\tilde{k}_2 +J,
k_3=\tilde{k}_3$, with tilde quantities kept fixed. The analysis finds the 
result
\be
< {\cal O} (x_1) {\cal O} (x_2) {\cal O}(x_3) > = \frac{J^{1-\tilde{k}_3/2}}{N}
\frac{\sqrt{\tilde{k}_1!\tilde{k}_2!\tilde{k}_3}}{\tilde{\alpha}_3!}
\frac{<\tilde{\cal C}^{I_1}\tilde{\cal C}^{I_2} \tilde{\cal C}^{I_3} >}
{|x_{12}|^{2(J+\tilde\alpha_3)}|x_{23}|^{2\tilde\alpha_1}
|x_{13}|^{2\tilde\alpha_2}}
\ee
which we can already see that is subleading with respect to
 the previous in the 
large J limit.

As we mentioned, we can take more easily the Penrose limit in the $(t,\vec{u})$
coordinates. Correspondingly, on the boundary we must make the conformal 
transformation + Wick rotation to the Lorentzian cylinder via $x_i= e^{\tau}
\hat{e}_i, \tau= it$. 

The conformal transformation acts as usual also on the boundary CFT 
operators, such that 
\be
{\cal O}'^I(\tau, \hat {e}_i)= e^{k\tau}{\cal O}^I(x)
\ee
and if we choose $-\tau_1 \gg 1$ ($x_1$ very close the origin so that 
we can say $|x_{12}|\sim |x_2|$),
the 2-point function becomes 
\be
< {\cal O}'^{I_2}{\cal O}'^{I_1}>=\delta^{I_1I_2} e^{-k(\tau_2-\tau_1)}
\ee
which means that the state (on the cylinder), which corresponds to the operator
${\cal O}(x)$ is ${\cal O}'^I(\tau, \hat{e})|0>= e^{k\tau}|I>$, where 
$<I_2|I_1>=\delta^{I_1I_2}$. 

For the 3-point function, \cite{mp} chose then 
to have $-\tau_1\gg 1$ also, 
which is to say $x_1$ was chosen as (very close to) 
the origin on the plane, which is always possible.

But they also chose in the case of the 3-point function 
$\tau _2 \gg 1$ ($x_2$ at infinity), which is not always possible, since 
we want to keep the metric of $S^3 \times R$ invariant. It is possible to do 
that by a conformal transformation, but that takes us away from the cylinder.

If one takes nevertheless also $\tau _2 \gg 1$, such that 
$|x_{12}|\simeq e^{\tau_2}\simeq|x_{23}|, |x_{13}|\simeq e^{\tau_3}
$, one then can put the SYM 3 point function in a matrix element form 
and get 
\be 
<I_2|{\cal O}^{I_3}(\tau_3, \hat{e}_3)|I_1> =\frac{J^{1-\tilde{k}_3/2}}{N}
e^{-\tau_3(k_1-k_2)}\frac{\sqrt{\tilde{k}_1!\tilde{k}_2!\tilde{k}_3}}{\tilde
{\alpha}_3!}<\tilde{\cal C}^{I_1}\tilde{\cal C}^{I_2}\tilde{\cal C}^{I_3}>
\ee
Again, this result is subleading (proportional to $J^{1-\tilde{k}_3/2}/N$),
but as mentioned it is put in a form that looks like a string matrix element.
One would think one could rescale ${\cal O}_3$ \cite{mp},
 but one can't do that, since it is already normalized!

Turning now to the AdS side, the large J limit becomes  
 the Penrose limit, taken  by setting  $t=x^+ +x^-/R^2, \psi = x^+
-x^-/R^2, \vec{u}=\vec{w}/R, \vec{v}=\vec{y}/R$.

If one hopes to have the same type of holography as in AdS, one takes the
Penrose limit of the bulk to boundary propagator in $(t,\vec{u})$ coordinates,
after which the propagator 
becomes independent of $\vec{u}$ (or $\vec{x}$, or rather $\vec{w}$, after 
the Penrose limit) as well as $\hat{e}'$, and instead becomes one dimensional 
dependent (on $x^+$ and $t'$)
\be
K(x^+, x^-, \vec{w}; t', \hat{e}')= \frac{[A(\Delta)]^{1/2}}{2^{\Delta}
cos^{\Delta}[(1-i\epsilon)(x^+-t')]}
\label{ppprop}
\ee
This is the propagator from the center of AdS (u=0) to the boundary 
($u=\infty$), in the pp limit.


We note here already the problem. The propagator used above connects 
the boundary with the center of AdS, but the Penrose limit focuses only 
on the center of AdS, so use of (\ref{ppprop}) should give a zero result.

If one nevertheless takes this propagator and calculates the string 
scattering, we will see that one still gets a result that is zero in 
the Penrose limit.

One can use the usual AdS-CFT prescription of Witten to relate the partition 
functions $Z_{SYM}=Z_{string}$, and as the string partition function can be 
expressed formally and schematically as
\be
Z= \int DX^{\mu}Dh_{\alpha\beta}... e^{iS} = 
\int DX^{\mu}Dh_{\alpha\beta}... e^{iS_0}(1+\frac{1}{2\pi}\int d^2\sigma
\sqrt{h}h^{\alpha\beta} \partial_{\alpha}X^{\mu}\partial_{\beta}X^{\nu}
f_{\mu\nu}(X^{\rho})+...)
\ee
where $f_{\mu\nu}(X^{\rho})$ is the graviton wave function, in this case
$h_{++}$, and the nontrivial insertion is the vertex operator for the 
graviton (however, that is not of definite momentum as usual). It is also 
equal to 
\be
S_{int}(h_{++}) =\frac{1}{4\pi \alpha '} \int_{-\infty}^{+\infty}
dx^+ \int_0^{2\pi \alpha ' |p_-|}d\sigma h_{++}(x^+,y)
\ee
and then 
\be
\frac{\delta Z}{\delta \phi_0}|_{string} = 
<I_2|\frac{\delta \;iS_{int}(h_{++})}{\delta \phi_0(x_3)}|I_1>
\ee
is equal to
\be
\frac{\delta Z}{\delta \phi_0}|_{SYM}= <I_2|{\cal O}_3(x_3)|I_1>|_{SYM}
\ee
In the string calculation, 
\be
\frac{\delta h_{++}}{\delta \phi_0(t', \hat{e}')}
\ee
is related via a rescaling to the bulk to boundary propagator (\ref{ppprop}).
The rescaling is performed to give the canonical scalar KK fields of
\cite{lmrs}, $s^I\sim (R^{3/2}/N) \phi^I$, after which the metric perturbations
corresponding to these scalars are calculated. They are $h\sim R^2 k s^I Y^I$
and the final formula is 
\be
\frac{\delta h_{++}}{\delta \phi_0(t', \hat{e}')}
= \frac{R^{2-k_I}(k_I+1)\sqrt{k_I}}{N 2^{k_I/2}cos ^{k_I+2}
[(1-i\epsilon )(x^+-t')]}{\cal C}^I(\vec{y})
\ee
(note that in $h_{MN}dx^Mdx^N$ only $h_{++}=h_{tt}+h_{\psi\psi}$ survives).
Here $k_I=\tilde{k}_3$ so is finite in the pp limit and $R^2\propto J$, so 
the string amplitude is 
$\propto J^{1-\tilde{k}_3/2}/N$, same as the gauge theory amplitude, 
thus it is not finite. 

The final result depends only on the boundary time $t_3$ (as the $\hat{e}_3$
dependence was lost in the pp limit), and matches the SYM result, however 
it can't be reexpressed only in pp wave (finite) quantities, since as we 
saw, it is actually subleading in the limit. One needs explicitly R and N.

So how could we salvage this calculation and get a finite amplitude in the 
pp wave limit? As we mentioned, we need to take 
${\cal O}_3$ to have large charge also. If we don't restrict to small charge,
\be
\frac{\delta h_{++}}{\delta \phi_0(t', \hat{e}')}
= \frac{4R^2}{k+1}(k^2+\partial_t^2)
\frac{\delta s^I}{\delta \phi_0(t', \hat{e}')} Y^I=
\frac{R^{2}(k_I+1)\sqrt{k_I}}{N 2^{k_I/2}cos ^{k_I+2}
[(1-i\epsilon )(x^+-t')]} Y^I
\ee
but now we don't have $Y^I=R^{-k} {\cal C}(\vec{y})$ but rather
\be
Y^I(\psi, \vec{v})= [\begin{pmatrix} k&\\  J&\end{pmatrix}]
^{1/2}2^{-J/2} e^{iJ\psi} (1-v^2)^{J/2}
\tilde{\cal C}^I(\vec{v})
\ee
and since $\tilde{\cal C}(\vec{v})=R^{-\tilde{k}}\tilde{\cal C}(\vec{y})$ 
we get 
\be
Y^{I_1}\simeq 2^{-k_1/2} \frac{(p^+)^{\tilde{k}_1/2}}{\sqrt{\tilde{k}_1!}}
f(x^+, x^-, \vec{y})
\ee
It is interesting to note that if we took the limit now on $h_{++}$, we would
get therefore
\be
\frac{\delta h_{++}}{\delta \phi_0(t', \hat{e}')}
\sim 2^{-J} \frac{J^{5/2}}{N}
\ee
and this is the wrong type of result: if we ignore the $2^{-J}$ as part of the 
space dependence, it is too big! ($J^{5/2}/N$ instead of $J^{3/2}/N$). 

The reason for this discrepancy is rather sneaky. By taking first $t'=-\infty$
before the Wick rotation and the pp limit, we saw that in the propagator
$2^{-k}cos^{-k}(...) $ was replaced by $e^{-ikt}(1+u^2)^{-k/2}$ 
(see Appendix A and specifically (\ref{propwick}) for details) and the 
interesting fact is that (apart from getting rid of the unwanted $2^{-k}$ 
factor) now, unlike before, $(k^2+\partial_t^2) \delta{s^I}/
\delta \phi_0\sim (k^2+\partial_t^2)K_{B\partial}=0$ in first order, 
so we have to look 
for subleading behaviour in J. 

Indeed, as \cite{mp} show, one gets
\be
\frac{\delta s^{I_1} }{\delta \phi_0}
 Y^{I_1} = \frac{|p^+|^{\tilde{k}_1/2}\sqrt{J}}{4N 
\sqrt{\tilde{k}_1!}}e^{-ip^+x^--i\tilde{k}_1x^+-|p^+|(w^2+y^2)/4}
\tilde{\cal C}^{I_1}(\vec{y})
\ee
and now the factors leading to $h_{++}$ ($4R^2/(k+1) (k^2+\partial_t^2)$)
don't bring an extra $J^2$ as before,
but rather J, leading to a good
\be
h_{++}\sim\frac{J^{3/2}}{N}
\ee
(finite) behaviour.

In conclusion, two things must be done in order to obtain something finite in 
the pp limit. The first is to take ${\cal O}_3$ to have dimension
 of order J as well
(formally, in \cite{mp}, one went from a $k_3=\tilde{k}_3$ perturbation to
a $k_1=J+\tilde{k}_1$ perturbation). That already forces us as far as leading 
behaviour goes to take extremal correlators (as the only ones that behave as
$J^{3/2}/N$). 

The second is to take propagators where the boundary point has been already 
put to the origin and use this new propagator.

But in going from the 
$h_{++}$ of \cite{mp} to the finite perturbation , there is a change of 
interpretation. The first used the bulk to boundary propagator
for $\delta h_{++}/\delta \phi_0(t_3, \hat{e}_3)$, namely
\be
<0|s^{I_3}(t, \vec{x}){\cal O}^{I_3}(t_3, \hat{e}_3)|0>
\ee
which contained an operator of finite $k_3$, therefore can't be mapped to 
a string state on the pp wave (would have zero $p^+$), and moreover it
still has dependence on the boundary point, therefore the role of $O^{I_3}$
is to couple to a boundary source, thus relating it to the 3-point 
SYM function $<I_1|{\cal O}(t_3, \hat{e}_3)|I_2>|_{SYM}$. 

But in our case, 
\be
<0|s^{I_1}(t, \vec{x}) {\cal O}^{I_1}(-\infty)|0>
\ee
is mapped to 
\be
<0|s^{I_1}|I_1>(x^+; |w|, x^-, \vec{y})
\ee
which has no more dependence on the boundary point, defines a state, and 
depends only on $x^+$ and coordinates transverse to the boundary.

It is conceivable therefore that there would be a way of constructing a 
vertex operator of definite momentum along $x^+$ and integrated as before over
the whole space. 

Unfortunately, it is not clear how to calculate the finite amplitude 
in this first quantized string formalism, since $p^+$ changes in the amplitude
($J_1=J_2+J_3$, thus $p^+_1=p^+_2+p^+_3$).
It is rather a string  field theory calculation (which was already done by 
many people) which makes more sense.

But one sees that however the calculation will be done, 
the result will be correct. In \cite{mp}, $h_{++}$ has 
already the final leading dependence, of $J^{1-\tilde{k}_3/2}/N$, since 
the evaluation of string oscillators brings only finite quantities, the same 
as the integrations. In our case, $h_{++}$ is already of order $J^{3/2}/N$, 
so one just needs the correct prescription for the vertices to get the right 
result.

\section{S matrices from SYM}

{\em Set-up}

The natural observables in AdS-CFT are correlators. But in flat space 
we have the LSZ formula that relates them to S matrices (observable), 
\bea
&& \prod_{i,j} \int d^4x_i e^{ip_ix_i} \int d^4 y_j e^{-ik_jy_j}
<\Omega|T\{ \phi(x_1)...\phi(x_n)\phi(y_1)...\phi(y_m)\}|\Omega>
\nonumber\\
&& \sim \lim_{p_i^0\rightarrow E_{p^i}, k_j^0\rightarrow E_{k_j}}
(\prod_i \frac{\sqrt{Z_i}i}{p_i^2-m_i^2+i\epsilon})
(\prod_j \frac{\sqrt{Z_j}i}{k_j^2-m^2+i\epsilon}) S(p_1,...p_n;k_1,...k_m)
\eea
where 
\be
\int d^4 x e^{ipx}<\Omega| T\{\phi(x)\phi(0)\}|\Omega>=\frac{iZ}{
p^2-m^2-i\epsilon}
\ee
So amputate the momentum space correlator near on shell 
and multiply by $\sqrt{Z}$'s and get the S matrix!

That implies a prescription for the AdS case as well. Indeed, as Giddings 
\cite{gid} notices, AdS-CFT takes the form
\be
<T({\cal O}(\vec{y}_1)...{\cal O}(\vec{y}_2))> = \int \Pi_i
[d^4 x_i K_F(\vec{y}_i, x_i)]G_T(x_1,...x_n)
\ee
where $K_F$ is the full multiloop bulk to boundary propagator and $G_T$
is the full amputated bulk n-point function. So if one could define the 
amputation process and the multiplication by $\sqrt{Z}$ on the boundary 
correlators, one would have an S matrix. But of course AdS doesn't admit 
an S matrix, so one has also to take a flat space (or pp wave!) limit. 

Moreover, one needs Lorentzian signature to define S matrices, so one 
needs to go to global AdS where that can be done (in the Poincare patch 
it is hard, since there we can't change the non-normalizable 
propagator to a normalizable one as we will do 
below for global coordinates). 

Then one has to define quantities which become in the flat space limit the 
5-momentum in Minkowski space. The natural candidate for the energy E is the 
conjugate to the global time $\tau$. So perform a Fourier transform and 
identify the variable $\omega $ with $ER$. From the point of the CFT, there 
remains a unit vector $\hat{e}$, which is however in position space. However, 
\cite{gid} argues  it should become the momentum unit vector in AdS. So
\be
k_{(5)}= E(1, \hat{e})
\ee
and so on-shell massless AdS momenta correspond to off-shell CFT momenta. 
But the discrete AdS states (normalizable modes) still correspond to the 
discrete states $(\Delta, n,l)$ 
on the cylinder ($ \omega= 2h_++2n +l$). Actually, we 
can also think of the cylinder states as on-shell, with a discrete 
set of masses. That is, $(\Delta, n=0, l)$ are states corresponding on the 
cylinder to operators on the plane of dimension $\Delta$, while n is an 
off-shell index which can be identified with ordering a tower $\{ \Delta_n 
\}_n$ of operators. Indeed, we will shortly see that n correponds in the 
pp wave limit to the radial oscillator number for the transverse (AdS)
oscillators, thus on SYM it is identified with the number of $\Phi_i$ 
insertions (non-U(1)-R charged scalars). Thus also in the flat space 
limit n indexes a tower of operators. 

Giddings only made this identification for massless states in AdS, but since 
we want to take the pp wave limit, we will need to extend it to
 massive states as well. In that case therefore, at large R, $2n/R= E-m$
and $\Delta/R= m$. We will see shortly that this gives the correct 
behaviour.

As \cite{gid} showed, in the large R limit, the bulk to bulk propagator 
of AdS space in global coordinates goes over to the flat space propagator.
For the bulk to boundary propagator, the discussion is a bit more involved,
but before it we need to study the flat space and pp wave
limits of wavefunctions. See also \cite{cd} for the pp wave limit of
supergravity couplings. 

\vspace{.5cm}

{\em Flat space and pp wave limits of wavefunctions}

We will look at 
the harmonic oscillator (the limit in which harmonic oscillator 
states go over to flat space free waves), since this will be needed to 
get flat space from the pp wave, then at AdS wavefunctions, both in the 
flat space and the pp wave limits.

Using the formulas derived in Appendix C, we obtain:

\begin{itemize}

\item The harmonic oscillator wavefunctions 
\be
\phi_n(x)=\frac{(m\omega)^{1/4}}{[\sqrt{\pi}2^nn!]^{1/2}}e^{-\rho^2/2}
H_n(\rho)
\ee
where $\rho = \sqrt{m\omega}x$ become in the large n limit (or 
flat space limit, $m\omega \rightarrow 0$)
\be
\lim_{n\rightarrow \infty} \phi_{2n}(x)=\frac{1}{\sqrt{\pi}}[\frac{4m\omega}
{n}]^{1/4}cos (2\sqrt{m\omega n} x)
\ee
which are normalized 
\be
\int_0^L \phi_{2n}(x)\phi^*_{2p}(x)= \delta_{np}
\ee
with the momentum 
\be
k_n=\sqrt{2m \omega n}=\frac{2\pi n}{L}
\ee
This calculation is the same one that one needs to prove that the pp wave
wavefunctions become free waves in flat space, with $m \omega$ replaced 
by $p^+$.

\item
In the case of AdS wavefunctions, the radial wavefunction is 
\be
\chi_{nl}(r)=A_{nl}(\cos(r/R))^{\Delta}(\sin(r/R) )^l
P_n^{l+d/2-1, \nu}(\cos (2r/R))
\ee
In the general case of the large n, large $\Delta$ limit (but small l),  
$\Delta/R\simeq m$, $2n/R\simeq E-m\equiv E'$ so $2r/R=E'r/n$ and 
$\nu \simeq  n (2m)/E'$. 
We want to obtain the flat space wavefunctions for massive modes, with 
$E= E'+m$ and $ |\vec{k}|= E'\sqrt{1+\beta}$
Then 
\be
\chi_{nl}(r)= A_{nl}(\frac{R}{r\sqrt{1+\beta}})^{d/2-1}(1+\beta)^{-l/2}
J_{l+d/2-1}(E'r\sqrt{1+\beta})
\ee
where $\beta=2m/E'$,  gives in the large n and large R limit the 
flat space wavefunctions in spherical coordinates
\be
\phi_{nl\vec{m}}(t, r, \hat{e})\rightarrow 
\sqrt{2E}(\frac{e^{-iEt}}{\sqrt{2E}})
\frac{J_{l+d/2-1}(rE'\sqrt{1+\beta})}{r^{d/2-1}}Y_{l\vec{m}}
(\hat{e})
\ee
That means that its Fourier transform will be 
\be
\sum_{l\vec{m}}Y_{l\vec{m}}(\hat{e}')\phi_{nl\vec{m}}(t,r,\hat{e})
\propto e^{-iEt}e^{iE'r\sqrt{1+\beta}\hat{e}\hat{e}'}= e^{-iEt+i\vec{k}\vec{x}}
\ee
and where $|\vec{k}|= \sqrt{E^2-m^2}=E'\sqrt{1+\beta}$, thus our 
identification of E and m was correct.

\item To get pp wave wavefunctions from AdS wavefunctions we fix n and 
l and take
$\Delta/R^2\simeq p^+$= fixed. Then we have 
\be
A_{nl}\rightarrow R^{l+1/2} (p^+)^{l/2+d/4}\sqrt{\frac{2n!}{\Gamma(n+l+
d/2)}};\;\;\; (\cos\frac{r}{R})^{\Delta}(\sin \frac{r}{R})^l\rightarrow
e^{-\frac{r^2p^+}{2}}(\frac{r}{R})^l
\ee 
and using the limit (\ref{pplimit}) we have 
\be
\lim _{R\rightarrow \infty} P_n^{l+d/2-1, p^+ R^2}(\cos(2r/R))
= L_n ^{l+d/2-1}(p^+ r^2)
\ee
and we get the wavefunctions for pp wave oscillators in spherical coordinates
(with $p^+=\mu \omega$) as obtained in (\ref{wavefct}).

\end{itemize}

\vspace{.5cm}

{\em Flat space vs. pp wave limits for S matrices}

We saw that in the global AdS parametrization, AdS-CFT relates the AdS 
energy $E$ with 
$\omega_{nl}/R$, the spatial momentum direction in $AdS_5$,
$\vec{k}/|\vec{k}|$ with $\hat{e}$, and the AdS mass $m$ with $\Delta/R$.
Of course that means that we have to define R and the large R limit from 
SYM, but 
from the above information we deduced $2n/R = E-m$, so large n is the 
same as large R.

Notice that the above identifications are 
 only true in the large n, large R limit. Otherwise,
\be
(\Delta-\frac{d}{2})^2=\frac{d^2}{4}+m^2R^2
\ee
and extremality of correlators, $\Delta_1=\sum_i \Delta_i$ means
\bea
&&m_1^2 R^2= (n-2)(n-1) \frac{d^2}{4} +\sum_i m_i^2 R^2+ 2\sum_{ij} 
\sqrt{\frac{d^2}{4}+m_i^2 R^2}\sqrt{\frac{d^2}{4} +m_j^2 R^2} 
\nonumber\\&& +(n-2) d 
\sum_i \sqrt{\frac{d^2}{4}+m_i^2 R^2}
\eea
and we see that we recover $m_1=\sum_i m_i$ in the large R limit. 

But we notice that when we say mass, we mean 5d supergravity mass, 
which is related to the 10d mass $M$ by
\be
m_{AdS}^2=M^2 +\tilde{k}_{S_5}^2
\ee

To obtain the pp wave limit, we need to boost in an $S_5$ direction, but 
in a precise way. 

If we rescale 
\be
\tilde{x}^-=\frac{x^-}{R};\;\tilde{x}^+=\frac{x^+}{R}\Rightarrow
2p'^-=\frac{\bar{\Delta}-J}{R};\; 2p'^+=\frac{\bar{\Delta}+J}{R};\;\; 
(2E=\frac{\bar{\Delta}}{R}
;\; 2p^{\psi}=\frac{J}{R})
\ee
and keep $p'^-,p'^+$ finite (or $E, p^{\psi}$ finite) in the large R limit, 
we get flat space. We need instead to rescale 
\be
\tilde{x}^-=\frac{x^-}{\mu R^2};\;\; \tilde{x}^+=\mu x^+\Rightarrow
2p^-=\mu (\bar{\Delta} -J);\; 2p^+=\frac{\bar{\Delta} +J}{\mu R^2}
\ee
We still have 
\be
E_{AdS}=\frac{\bar{\Delta}}{R},\;\;\; p_{S_5}^{\psi}= \frac{J}{R}
\ee
but they are not finite.

Notice that in both cases we get a finite 
\be
E^2-p_{\psi}^2=
(2p^-)(2p^+)= (2p'^-)(2p'^+)=\frac{\bar{\Delta}^2-J^2}{R^2}= M^2
+\vec{p}_{transv}^2= M^2+\vec{k}^2_{(4), AdS}+\vec{k}^2_{(4),S}
\ee

We have written $\bar{\Delta}$ instead of $\Delta$ since there is 
a change of interpretation. Here 
\be
\bar{\Delta}=\omega_{nl}= \Delta +2n+l
\ee
takes into account both the ``off-shell index'' n (radial
 AdS oscillator number 
for the pp wave case) and the angular momentum l, i.e. AdS directions
oscillator numbers in spherical coordinates (as we saw). 
On the SYM $R^4$ plane, going from an operator ${\cal O}$ to ${\cal O}$ with 
insertions of $D_{(\mu_1}...D_{\mu_l)}-traces$ is equivalent on the cylinder 
$S^3\times R_t$ to taking the l-th KK mode on $S^3$, i.e. one with 
spherical harmonic $Y_{l\vec{m}}$. We could have derivatives contracted 
with each other, and $2n$ corresponds to the number of contracted derivative
insertions ($\sum_i N_{0i}=2n+l$ in the pp wave case).
Thus $\bar{\Delta}$ corresponds to taking 
also possible derivatives into account when counting
 the dimension of operators.

The momentum on $S_5$ is characterized in AdS by the spherical harmonic 
$Y_{l\vec{m}}(\tilde{\hat{e}})$, and it corresponds in SYM on the 
cylinder (for global AdS) also to the $(l,\vec{m})$ representation
of operators. For Poincare AdS, the $S_5$ momentum is determined in SYM 
by the number of $\Phi^i$ insertions (the same way as $Y_{lm}(\hat{e})$
corresponds to $D_{\mu} Z$ insertions), if they are in a large number 
(comparable to J). 

In the flat space limit, spherical harmonics become free waves. For example,
the ``spherical harmonic'' on a circle becomes
\be
Y_J= e^{iJ\phi}= e^{\frac{iJ y}{R}}\rightarrow e^{iky}; k\equiv \frac{J}{R}
\ee
For the 2-sphere we have $Y_{lm}(\theta,\phi)$ and large $m$ is as before,
whereas large $l$ is similar, as $P_l(\cos\theta)$ becomes $\cos k x$. So 
in general, we can say that $Y_{J\vec{m}}(\tilde{\hat{e}})\rightarrow 
e^{i\tilde{\vec{k}}\cdot \vec{x}}$, thus the $S^5$ momentum is determined 
by $Y_{J\vec{m}}$, i.e. operators that are in a representation that 
corresponds to $Y_{J\vec{m}}$ will have momentum $\vec{k}$.

\begin{itemize}

\item 
In conclusion, if we want to take the flat space limit directly, we take 
fixed $E_{AdS}=\bar{\Delta}/R$. For $\Delta\sim 1$, $n$ large, we get 
$m_{AdS}=0$ S matrices \cite{gid}. For $m_{AdS}=\Delta/R$ fixed, we 
get nontrivial mass and/or momenta. 
The sphere momentum $\vec{k}^2_{S_5}$ is defined 
by large $J\sim R\sim (g^2_{YM}N)^{1/4}$ giving fixed $p^{\psi}$,
 and maybe large number of $\Phi$ 
insertions (giving extra momenta on $S^5$, in the directions perpendicular
to $\psi$), $N_{\Phi_i}\sim (g^2N)^{1/4}$. 
The 10d mass M is obtained by the phases 
$e^{i\frac{2\pi n l }{J}}$ in the BMN operators \cite{bmn}, i.e. by having 
operators that have BMN phases and insertions on top of any representation 
corresponding to $Y_{J\vec{m}}$. 

\item
If we want instead to get the pp wave limit, we have to keep fixed $
\Delta /R^2$. But
\bea
&&E_{AdS}^2-\vec{k}_{4,AdS}^2= m^2_{AdS}=p_{\psi}^2+\vec{k}^2_{4,S}+M^2
\Rightarrow\nonumber\\&&
E^2=\frac{\Delta^2}{R^2}+2\frac{\Delta}{R^2}(2n+l)+\frac{(2n+l)^2}{R^2}
= m_{AdS}^2+
2\frac{\Delta}{R^2}(2n+l)+\frac{(2n+l)^2}{R^2}
\eea
So $n, l\sim1$ in order to have finite $\vec{k}^2_{4,AdS}= 2\Delta/R^2 (2n+l)
$. Thus $\Delta\sim R^2, n\sim 1, l\sim 1$.
The sphere momentum $\vec{k}^2_{S_5}$ is defined by 
$J\sim R^2\sim (g^2_{YM}N)^{1/2}$ and by small number of $\Phi $ insertions
(of order 1), that give the extra (discrete) momenta $N_{0i}$. Again, the 
 10d mass M is obtained by the phases 
$e^{i\frac{2\pi n l }{J}}$ in the BMN operators.

\end{itemize}

The difference between the pp wave limit and the flat space limit can 
be understood by looking at energy of string states in the pp wave,
\be
-2p^-=\mu \sum_i N_{0i}+\mu \sum_i \sum_{n_i}\sqrt{1+\frac{g_{YM}^2N}{J^2}}
= \sum_i \frac{p_i^2}{p^+}+\frac{M^2}{p^+}
\ee
If $N_{0i}\sim 1$, we get finite discrete $p_i$'s (``momenta''), and if $J\sim
g^2_{YM}N$ we get finite mass $M_{\{n_i\}}$.

If on the other hand, $J\sim (g^2_{YM}N)^{1/4}$, we get  $p^+\sim 1/R\sim
(g^2_{YM}N)^{-1/4}$, or rather, we have to redefine momenta to get finite 
results:
\be
-2p'^-=\frac{2p^-}{\mu (g^2_{YM}N)^{1/4}}=\sum_i\frac{N_{0i}}{
(g^2_{YM}N)^{1/4}}+\sum_i\sum_{n_i}\frac{(g^2_{YM}N)^{1/4}}{J}n_i
\ee
and again we get discrete mass and discrete momenta, if $N_{0i}\sim 
(g^2_{YM}N)^{1/4}$.

We have seen that the ``off-shell index'' n correponds after the pp 
wave or flat space limit to the radial oscillator number in the AdS 
directions (n and l are the spherical coordinate representation of 
the isotropic d-dimensional oscillator). 

Since now we want a spacetime interpretation, for the S matrix,  we need 
n since an $AdS_5$ on-shell state corresponds to an $S^3\times
R_t$ ``off-shell'' state (there is one extra degree of freedom, corresponding 
to the radial dimension). But we see that really the ``off-shell'' 
states are just reorderings of the on-shell states into a tower.

\vspace{.5cm}

{\em S matrix definition}
 
In order to extract the S matrix we have to amputate external 
legs. Usually, the momentum space correlators have poles when the momenta 
are on-shell. But now as we saw, the bulk to boundary AdS propagator 
has poles when the external frequencies are normalizable (when CFT states 
are ``on-shell'').

So \cite{gid} devises a trick (that doesn't seem to work in Poincare 
coordinates) that converts the bulk to boundary propagators $K_{B\partial}$,
which are (both in Poincare and in global 
coordinates) non-normalizable wavefunctions as we saw, into normalizable 
wavefunctions, thus defining the S matrix. One makes a time 
Fourier transform on the external legs of the AdS correlators, thus 
performing the same on the boundary leg of the 
bulk to boundary propagators, to $K_{B\partial} (\omega, \hat{e}; x')$.

Then one isolates one of the poles at external normalizable frequencies
by first going to $l,\vec{m}$ space by integrating with $Y_{l\vec{m}}(\hat{e})
$ and then defining
\be
\hat{K}(n,l,\vec{m}, x)=\lim_{\omega\rightarrow \omega_{nl}}K_{B\partial}
(\omega, l, \vec{m},x)= -2\nu R^{d-1}e^{i\omega_{nl}\tau}
k_{nl}\phi^*_{nl\vec{m}}
\ee
which is thus just a normalizable wavefunction,
and finally going back to position space on the boundary leg by 
\be
\hat{K}(E, \hat{e}; x)= \sum_{l\vec{m}}Y_{l\vec{m}}(\hat{e}) \hat{K}(n, l , 
\vec{m}, x)
\ee
He showed then that this becomes the free (normalizable) wavefunction
\be
K(E, \hat{e}; x)\rightarrow \frac{C(E,R)}{(2\pi)^{d/2}}e^{i\vec{k}\vec{x}}
\ee

Thus in SYM the procedure corresponds to acting on the r-point correlator 
with
\be
\prod_{j=1}^r(\sum_{l'_j\vec{m}'_j}Y_{l'_j\vec{m}'_j}(\hat{e}'_j))
\prod_{i=1}^r (\lim _{p_i\rightarrow\omega_{n'_il'_i}}
(p_i^2-\omega_{n'_il'_i}^2)\int d\hat{e}_i Y^*_{l'_i\vec{m}'_i}(\hat{e}_i))
\ee

However, as we just saw, the flat space limit works for the massive case 
as well, and we saw that the Fourier transform of the wavefunction is 
a free wave $e^{ikx}$ even then. 

To get the pp wave S matrix  we apply the same procedure, but keeping
$\Delta/R^2, n, l$ fixed. The wavefunctions work again, and we obtain 
the pp wave wavefunction, thus applying the procedure to SYM correlators
for the previous limit generates pp wave S matrices.
We also have to get an appropriate limit for the representation of 
the operators in the correlator (basically, take BMN operators).

\section{(Extremal) Correlators and AdS S matrices}

We will now analyze the correlators and see whether we can test this 
assumption. The pp wave (massive) case is quite involved, so we will 
study the massless case (flat space) of Giddings first, looking to apply 
the procedure to nonrenormalized correlators, so that we can test it.
As we mentioned, these cases are the 2 and 3-point functions and the 
extremal correlators. The general extremal correlators are also 
the only ones that remain in the pp wave limit (meaning that they are 
related to S matrix observables). We will analyze them afterwards.

{\em 2- and 3-point functions}

The momentum space free scalar two point function is 
\be
\int d^4x d^4 y e^{iky+ipx}\frac{1}{|x-y|^2}= \int d^4y e^{iy(k+p)}
\int d^4 x' \frac{e^{ipx'}}{x'^2}= \delta^4(k+p)\frac{1}{p^2}
\ee
One could also put y to zero by translational invariance, then obtaining 
the two point function minus the momentum conservation
\be
\int d^4 x \frac{e^{ipx}}{x^2}= \frac{1}{p^2}
\ee
For the 3-point function,
\be
f(x_1,x_2,x_3)= \frac{a_{123}}{x_{12}^ax_{13}^bx_{23}^c}
\ee
we have
\bea
&& f(p_1,p_2,p_3)= \int d^4x_1d^4x_2d^4x_3e^{i(p_1x_1+p_2x_2+p_3x_3)}
f(x_1, x_2, x_3)\nonumber\\&&
= \int d^4x_1 e^{ix_1(p_1+p_2+p_3)}
\int d^4 x_{21}d^4x_{31}e^{i(p_2x_{21}+p_3x_{31})}f(x_{21}, x_{31})
\nonumber\\
&&= \delta^4(p_1+p_2+p_3)a_{123}\int d^4xd^4y \frac{e^{ip_2x+ip_3y}}
{x^ay^b |x-y|^c}\nonumber\\&&
= \delta^4(p_1+p_2+p_3)a_{123}f(|p_2|, |p_3|,
\hat{e}_2\hat{e}_3 )
\eea
and again we get the overall delta function. It is however not a good idea
to fix translational invariance in our case
by putting one of the points to zero, since 
we need to use the Giddings procedure to extract S matrices.

Moreover now, for the global coodinate calculation in AdS, we need only 
to Fourier transform the SYM correlators in radial time.
At the begining of the previous section, we also put one of the 
points near the origin (negative infinite radial time). 
If we don't put it to zero, the x-space SYM 2-point function corresponding
to  AdS global 
coordinates (i.e., on the cylinder) is now 
\be
\frac{e^{ik(t_1+t_2)}}{|x_1-x_2|^{2k}}= \frac{1}{2^k}\frac{1}{(
\cos (t_1-t_2)- \hat{e}_1\hat{e}_2)^k}
\label{equ}
\ee
But \cite{gid} showed that that expression is also equal to 
\be
K_B(x,x')= c \int \frac{d\omega}{2\pi}\sum_{nl\vec{m}}
e^{i\omega(t-t')}\frac{k_{nl}^2Y_{l\vec{m}}^*(\hat{e})Y_{l\vec{m}}(\hat{e}')}
{\omega_{nl}^2-\omega^2-i\epsilon}
\label{bdprop}
\ee
where $\omega_{nl}=2k+2n+l$, and is actually valid even if $2k$ is not
integer and is $=\Delta$. That means that the momentum space SYM 2-point 
function is 
\be
<{\cal O}^{I_1}(p_1, \hat{e}_1); {\cal O}^{I_2}(p_2, \hat{e}_2)>=
c\delta^{I_1I_2}\delta(p_1+p_2)
\sum_{nl\vec{m}}\frac{k_{nl}^2Y^*_{l\vec{m}}(\hat{e}_1)Y_{l\vec{m}}(\hat{e}_2)
}{\omega_{nl}^2-p_2^2-i\epsilon}
\ee
Incidentally, that also identifies n as an off-shell index. Indeed, from the 
above one sees that the states of the CFT on the cylinder are labelled by 
$\Delta $ and l, but also by this off-shell index n. At the free level, when 
$\Delta =2k$, we can identify n with k, considered as indexing a tower of 
operators. At the interacting level, even n becomes non-integer, so it 
could still be indentified with the index of a tower of $\Delta_n$.

Let us now look at the nonextremal (general) SYM 3-point function on the 
cylinder. The exact result is written in Appendix D.

Let us understand how does the summation and integration over one set 
of variables dissappear when $\alpha =0$, i.e. for 
extremality. We see from (\ref{equ}) and 
(\ref{bdprop}) that $\alpha =0$ implies one should have 
\be
1= -\int \frac{d\omega}{2\pi}\sum_{nl\vec{m}}
e^{i\omega t}\frac{k_{nl}^2Y_{l\vec{m}}^*(\hat{e})Y_{l\vec{m}}(\hat{e}')}
{\omega^2-(2n +l)^2 -i\epsilon}
\ee
From the expression for $k_{nl}$ (\ref{knl}) we see that when $\alpha=0$ 
and $d=4$ we have a $\Gamma(-1)^2$ in the denominator, meaning that the 
coefficient is zero unless it's compensated. The integral over $\omega$ gives
$1/2\omega_{nl}= 1/2(2n+l)$. If we put n=0 we get then a compensating infinity
and obtain
\be
-\sum_{l\vec{m}}\frac{1}{\Gamma(-1)^2}\frac{\Gamma(-1)\Gamma(l)}{0!\Gamma(l+2)}
Y^*_{l\vec{m}}(\hat{e})Y_{l\vec{m}}(\hat{e}')=-\frac{\Gamma(0)}{\Gamma(-1)}
Y^*_{00}(\hat{e}) Y_{00}(\hat{e}')=1
\ee
where in the last line we had to select $l=0$ as well. 

This trivial calculation is of importance since that is what happens when 
we have extremal correlators: one of the $\alpha$ s becomes zero, and 
then the variables get reduced. Indeed, from the above simple calculation, 
we see that when $\alpha_1=0$, we select $n_1=l_1=0$ and then the 
only nonzero term in the square brackets in (\ref{three}) is the middle one,
one puts $\omega_{n_1l_1}=0=n_1=l_1$ and recovers the extremal 3-point 
function result. Since we might need to make an analytic continuation to 
reach the extremal case, it is important to see the correct limit. (in 
Poincare coordinates, there was a continuation involved in getting the 
3-point function, naively the result was zero in AdS, but there was an 
infinite integral coming from volume).

To deduce the S matrix a la Giddings we have now to act on the 3-point function
with 
\be
\prod_{j=1}^3(\sum_{l'_j\vec{m}'_j}Y_{l'_j\vec{m}'_j}(\hat{e}'_j))
\prod_{i=1}^3 (\lim _{p_i\rightarrow\omega_{n'_il'_i}}
(p_i^2-\omega_{n'_il'_i}^2)\int d\hat{e}_i Y^*_{l'_i\vec{m}'_i}(\hat{e}_i))
\label{giddi}
\ee
We know that in that case on the AdS side we just get a momentum 
delta function for the 3-point S matrix, and in general we expect a momentum 
conservation anyway. In the 3-point case the delta function comes from the 
integrals of free waves $e^{ikx}$, and we generalize this representation 
of the delta function to higher n-point function
cases, even though then the AdS calculation 
will be more involved. But we expect the S matrix to have an overall momentum 
conservation anyway. We will suppress the overall energy conservation 
which is obtained anyway, $\delta (p_1+p_2+p_3)$ or in general $\delta(
p_1+p_2+...+p_n)$.

In Appendix B we have shown how to write the delta function in a form 
appropriate for comparison with the SYM calculation.

\vspace{.5cm}

{\em Extremal correlators}

The statement (conjecture?) about extremal correlators 
is that they do not receive corrections (as for the general 
3-point function). The case generally treated is $k_1=k_2+...+k_n$, since 
it's easier to analyze. Then
\be
<{\cal O}^*(x_1){\cal O}(x_2)
 ...{\cal O}(x_n)>= \frac{a_{12...n}}{|x_{12}|^{2k_2}...|x_{1n}|^{2k_n}}
\ee
In the general extremal case ($k_1+...+k_n=k_{n+1}+...+k_{n+m}$), the powers 
are more complicated. Note that if we put (by translational invariance)
 $x_1=0$ and go to $S^3\times R$ and rotate to Lorentzian signature we get 
\be
<{\cal O }^*(t_1, \hat{e}_1){\cal O}(t_2, \hat{e}_2)
...{\cal O}(t_n, \hat{e}_n)>=a_{1...n}e^{ik_1t_1-ik_2t_2-...-ik_n t_n}
\ee
For the general extremal correlator $k_1+..+k_n=\bar{k}_1+
...+\bar{k}_m$ in the special case $|x_{1i}|\ll |x_{ij}|, i=1,n; j=1,m$
the powers (approximately) combine to give a similar answer:
\be
<{\cal O}^*(t_1, \hat{e}_1...{\cal O}^*(t_n, \hat{e}_n){\cal O}(\bar{t}_1,
\hat{\bar{e}}_1)....{\cal O}(\bar{t}_m, \hat{\bar{t}}_m)>
=a_{1...n;1...m}e^{ik_1t_1+...+ik_nt_n-i\bar{k}_1\bar{t}_1-...-i\bar{k}_m
\bar{t}_m}
\ee
However, note that this answer is independent of $\hat{e}_i$, so cannot be 
correct. The point is that if we Fourier transform over the whole plane, we
have the option of putting one coordinate to zero as we saw above for the 
3-point function (all we miss is the overall delta function), or otherwise 
it gets shifted away anyway when integrating. But if we just Fourier 
transform over the energy, we can't. 

Let us look at the extremal 3-point function ($k_1=k_2+k_3$)
without fixing any point. The 3-point 
function on the Lorentzian cylinder, Fourier transformed in energy is 
\bea
&&f_{123}(p_1, \hat{e}_1;p_2, \hat{e}_2; p_3, \hat{e}_3)
= a_{123}\delta(p_1+p_2+p_3)
\int dt_2' \frac{e^{it_2'(p_2-k_2)}}{(1+e^{-2it_2'}-2e^{-it_2'}
\hat{e}_1\hat{e}_2)^{k_2}}
\nonumber\\&&
\int dt_3' \frac{e^{it_3'(p_3- k_3)}}{(1+e^{-2it_3'}-2e^{-it_3'}
\hat{e}_1\hat{e}_3)^{k_3}}
\eea
here as before $t_2'=t_{21}, t_3'=t_{31}$ and we have the product of two 
free 2-point functions:
\bea
&&
f_{123}(p_1, \hat{e}_1;p_2, \hat{e}_2; p_3, \hat{e}_3)
= a_{123}\delta(p_1+p_2+p_3)
\sum_{n_2l_2\vec{m}_2}\frac{k_{n_2l_2}^2Y^*_{l_2\vec{m}_2}(\hat{e}_1)
Y_{l_2\vec{m}_2}(\hat{e}_2)
}{\omega_{n_2l_2}^2-p_2^2-i\epsilon}
\nonumber\\&&
\sum_{n_3l_3\vec{m}_3}\frac{k_{n_3l_3}^2Y^*_{l_3\vec{m}_3}(\hat{e}_1)
Y_{l_3\vec{m}_3}(\hat{e}_3)
}{\omega_{n_3l_3}^2-p_3^2-i\epsilon}
\eea
And for the general extremal n-point function we similarly get 
\bea
&&
f_{12..r}(p_1, \hat{e}_1;p_2, \hat{e}_2;...; p_r, \hat{e}_r)
= a_{12...r}\delta(p_1+p_2+...+p_r)
\sum_{n_2l_2\vec{m}_2}\frac{k_{n_2l_2}^2Y^*_{l_2\vec{m}_2}(\hat{e}_1)
Y_{l_2\vec{m}_2}(\hat{e}_2)
}{\omega_{n_2l_2}^2-p_2^2-i\epsilon}...
\nonumber\\&&
\sum_{n_rl_r\vec{m}_r}\frac{k_{n_rl_r}^2Y^*_{l_r\vec{m}_r}(\hat{e}_1)
Y_{l_r\vec{m}_r}(\hat{e}_r)
}{\omega_{n_rl_r}^2-p_r^2-i\epsilon}
\eea
Now we will apply the Giddings procedure \cite{gid}
on the general 3-point function, and then particularize for the extremal 
correlators.
Doing the spherical harmonic integrals we get 
\bea
&&\sum_{m_1m_2m_3}\frac{1}{(4\pi)^{3/2}}(2l_1+1)(2l_2+1)(2l_3+1)
\sqrt{(2l_1'+1)(2l_2'+1)(2l_3'+1)}(l_1 l_3 l_2'; m_1 m_3 m_2') 
\nonumber\\ && (l_1 l_3 l_2' ;0 0 0 ) 
(l_1 l_2 l_3'; m_1 m_2 m_3')(l_1 l_2 l_3' ; 0 0 0) (l_2 l_3 l_1' ; m_2 m_3 
m_1') (l_2 l_3 l_1 '; 0 0 0)
\eea
Then multiplying with $p_i^2-\omega_{n_il_i}^2$ and taking the limit we 
pick out only certain terms in the sum over n,l: $l_2+l_3=l_1', 
l_3-l_1=l_2', l_1+l_2=l_3'$ (actually, only 2n+l is defined). 
Then, when taking the sum with 
\be
\sum_{l_i' m_i'}Y_{l_1' m_1'}(\hat{e}_1)Y_{l_2'm_2'}(\hat{e}_2)
Y_{l_3'm_3'}(\hat{e}_3)
\ee
we have fixed $l$ s in terms of $l'$ s, so we can't use the Gaunt formula 
in reverse! Notice though that one still remains with an unsaturated
 zero, coming from $p_1^2-\omega_{n_1l_1}^2$ (the rest cancel against the 
poles in the 3-point function).

In the extremal case we get the spherical harmonic integrals
($l_1=m_1=0$)

\bea
&&\int d\hat{e}_1 Y^*_{l_2m_2}(\hat{e}_1)Y^*_{l_3m_3}(\hat{e}_1)
Y^*_{l_1'm_1'}(\hat{e}_1)\int d\hat{e}_2 Y_{l_2m_2} (\hat{e}_3)Y^*_{l_3'm_3'}
(\hat{e}_3)\int d\hat{e}_2 Y_{l_3m_3}(\hat{e}_2)Y^*_{l_2'm_2'}(\hat{e}_2) 
\nonumber\\&&
=\frac{1}{\sqrt{4\pi}}\sqrt{(2l_2+1)(2l_3+1)(2l_1'+1)}(l_2 l_3 l_1'; m_2 m_3
m_1')^*(l_2 l_3 l_1' ; 0 0 0 )^*\delta_{l_2 l_3'}\nonumber\\&&
\delta_{m_2m_3'}
\delta_{l_3l_2'}\delta_{m_3 m_2'}
\eea
which easily generalizes to (renaming $l_2$ as $l_3$ and viceversa, in order
to generalize)
\be
I_{\{ l_i m_i \} l_1' m_1'}^* \delta_{l_2l_2'}\delta_{m_2m_2'}...\delta_{l_n
l_n'}\delta_{m_n m_n'}
\ee
So that the final result for the extremal S matrix from SYM is 
\bea
&&f_{12..r}(p_1, \hat{e}_1;p_2, \hat{e}_2;...; p_r, \hat{e}_r)
= a_{12...r}\delta(p_1+p_2+...+p_r) 
\nonumber\\&&\sum_{\{ l_i' m_i' \} }
k_{n_2'l_2'}^2...k_{n_r'l_r'}^2
(p_1^2-\omega_{n_1'l_1'}^2)I_{\{ l_i', m_i'\}l_1'm_1'}
Y_{l_1'm_1'}(\hat{e}_1)Y_{l_2'm_2'}(\hat{e}_2)
...Y_{l_r'm_r'}(\hat{e}_r)
\eea
which begins to look like (\ref{deltasph}) (we have both the spherical 
harmonics and the $I_{l_im_i}$ term). 

But we still have two operations to perform: to multiply with the coefficient 
giving the AdS 3-point function and to multiply with $C^{-1}(E_i, R)$. 

The action
\be
S=\frac{1}{2}\int \sum_I\eta_I[(\partial\phi_I)^2 +m^2\phi_I^2]+\lambda\int 
\phi_1\phi_2\phi_3
\ee
gives the two point function (see \cite{lmrs} and \cite{fmmr})
\be
<{\cal O}{\cal O}>=\eta \frac{\Gamma(\Delta+1)}{\pi^{d/2}\Gamma(\Delta-d/2)}
\frac{2\Delta-d}{\Delta}\frac{1}{|x-y|^{2\Delta}}
\equiv\frac{\eta f}{|x-y|^{2\Delta}}
\ee
and the 3-point function
\bea
&&<{\cal O}_1{\cal O}_2{\cal O}_3> = -\frac{\lambda}{|x-y|^{2\alpha_3}
|y-z|^{2\alpha_1}|z-x|^{2\alpha_2}}\frac{\Gamma(\alpha_1)\Gamma(\alpha_2)
\Gamma(\alpha_3)}{2\pi^d\Gamma(\Delta_1-d/2)\Gamma(\Delta_2-d/2)\Gamma(\Delta
_3-d/2)} \nonumber\\&&
\Gamma(\frac{\Delta_1+\Delta_2+\Delta_3-d}{2})
\equiv -\frac{\lambda b}{|x-y|^{2\alpha_3}
|y-z|^{2\alpha_1}|z-x|^{2\alpha_2}}
\eea
It is not clear by what coefficient should we multiply in the general 
extremal case, since then it is not even clear what calculation one should 
do in AdS. 

For the extremal 3-point function though, if $a_{123}$ is the coefficient 
of the normalized 3-point function, then 
\be
a_{123}=\frac{\lambda b}{\sqrt{\eta_1f_1\eta_2f_2\eta_3f_3}}
\ee
but for the S matrix we want to look only at the AdS 3-point function for 
$\lambda =1$, i.e at the b coefficient, whereas in SYM we get the $a_{123}$ 
coefficient, so we should multiply the SYM result by $b/a_{123}$. Putting 
also the $C^{-1}(E_i, R)$ factors we get for the 3-point S matrix 
\bea
&&f_{123}(p_1, \hat{e}_1;p_2, \hat{e}_2; p_3, \hat{e}_3)
= \delta(p_1+p_2+p_3) \frac{\Gamma(\alpha_1)\Gamma(\alpha_2)
\Gamma(\alpha_3)}{2\pi^d\Gamma(\Delta_1-d/2)\Gamma(\Delta_2-d/2)\Gamma(\Delta
_3-d/2)} \nonumber\\&&
\Gamma(\frac{\Delta_1+\Delta_2+\Delta_3-d}{2})
\sum_{\{ l_i' m_i' \} }
k_{n_2'l_2'}^2k_{n_3'l_3'}^2 C^{-1}_1C^{-1}_2C^{-1}_3
(p_1^2-\omega_{n_1'l_1'}^2)\nonumber\\&&
I_{\{ l_i', m_i'\}l_1'm_1'}
Y_{l_1'm_1'}(\hat{e}_1)Y_{l_2'm_2'}(\hat{e}_2)
Y_{l_3'm_3'}(\hat{e}_3)
\eea
In general be will have a factor analog to b, which presumably 
will also have a divergent $\Gamma(\alpha)$, so 
\bea
&&f_{12..r}(p_1, \hat{e}_1;p_2, \hat{e}_2;...; p_r, \hat{e}_r)
= b \delta(p_1+p_2+...+p_r) 
\nonumber\\&&\sum_{\{ l_i' m_i' \} }
k_{n_2'l_2'}^2...k_{n_r'l_r'}^2 C^{-1}_1C^{-1}_2...C^{-1}_r
(p_1^2-\omega_{n_1'l_1'}^2)I_{\{ l_i', m_i'\}l_1'm_1'}
Y_{l_1'm_1'}(\hat{e}_1)Y_{l_2'm_2'}(\hat{e}_2)
...Y_{l_r'm_r'}(\hat{e}_r)
\eea
At large R,
\bea
&& C\sim \frac{2^{2-\nu}}{\Gamma(\nu)}(-1)^{ER/2-h_+}(ER)^{\Delta}
\sim \frac{2^{2-\Delta +d/2}}{\Gamma(\Delta -d/2)}(-)^{n+l/2}(2n)^{\Delta}
\nonumber\\&&
k_{nl}^2\sim \frac{2ER}{\Gamma(\nu+1)^2R^{d-1}}(\frac{ER}{2})^{2\nu}
\sim \frac{2^2}{\Gamma(\Delta-d/2+1)^2R^{d-1}}n^{2\Delta -d+1}
\eea
where the second expression as in terms of SYM parameters. 

Then, using also $\Delta_2+\Delta_3=\Delta_1$ we get 
\bea
&&k_2^2k_3^2C_1^{-1}C_2^{-1}C_3^{-1}\sim (\frac{n_2}{n_1})^{\Delta_2+1-d}
(\frac{n_3}{n_1})^{\Delta_3+1-d}\frac{2^{-3d/2-2}}{R^{2(d-1)}n_1^{2(d-1)}}
\nonumber\\&&
\frac{1}{(\Delta_2-d/2)^2(\Delta_3-d/2)^2}\frac{\Gamma(\Delta_1-d/2)}{
\Gamma(\Delta_2-d/2)\Gamma(\Delta_3-d/2)}
\eea
Now we also see how the zero in $p_1^2-\omega_1^2$ is cancelled, since 
\be
p_1^2-\omega_1^2\equiv p_1^2-(\omega_2+\omega_3)^2\sim 2p_1R(\Delta_2+
\Delta_3-\Delta_1)= 4p_1R\alpha_1
\ee
and that gets cancelled because $\Gamma(\alpha_1)\alpha_1=1$.

Putting everything together we get
\bea
&&f_{123}(p_1, \hat{e}_1;p_2, \hat{e}_2; p_3, \hat{e}_3)
= \delta(p_1+p_2+p_3)\frac{(p_2/p_1)^{\Delta_2+1-d}(p_3/p_1)^{\Delta_3+1-d}}
{p_1^{2d-3}}\nonumber\\&&
\frac{2^{d/2-3}}{\pi^dR^{4d-5}}\frac{\Gamma(\Delta_2)\Gamma(\Delta_3)}{
\Gamma(\Delta_2-d/2+1)^2\Gamma(\Delta_3-d/2+1)^2}
\Gamma(\Delta_2+\Delta_3-\frac{d}{2})
\nonumber\\&&
\sum_{\{ l_i' m_i' \} }
I_{\{ l_i', m_i'\}l_1'm_1'}
Y_{l_1'm_1'}(\hat{e}_1)Y_{l_2'm_2'}(\hat{e}_2)
Y_{l_3'm_3'}(\hat{e}_3)
\eea

Except for the numerical factors which are different, the angular and 
momentum dependence is the same as in (\ref{deltasph}),
 except for an extra factor of $p_1$ in the 
denominator, and if we fix the constants $c=3n/2-2+\sum_i b_i$ and 
put $0=d-5/2-
\Delta_i$. So we still need to find a procedure to somehow get rid of the 
unwanted factors (in d=4) of $(p_2/p_1)^{\Delta_2 -3/2} (p_3/p_1)^{\Delta_3
-3/2}$, and of the gamma functions containg numerical $\Delta$ factors. 
Conceivably, there should be a procedure that renormalizes the external 
legs such that we get the correct S matrix, as the angular momentum dependence
was correct. Or maybe the problem is the fact that loop corrections 
modify the poles of the external propagators
(as suggested by Giddings that could happen), and so maybe the residues are 
also modified, this giving the discrepancy that we found. 

One should really analyze the massive and pp waves 
cases and obtain the massless flat space case as a limit. 

\section{Conclusions}

In this paper we have analyzed the possibility to obtain S matrices on 
flat space and pp waves from SYM. The question of pp waves is of interest 
in several respects. First, it is a nontrivial gravitational background, 
and second, we have seen that the pp wave limit already focuses in on a 
geodesic in the middle of AdS, so the hardest problem in the case of the 
flat space limit (getting rid of the boundary contributions) 
seems already solved. Of course, we need to make sure that we are indeed
in the pp wave limit. 

For that, we have looked again at the argument that only extremal correlators 
 survive the pp wave limit (in the same way that only large R charge operators
survive it), at least as far as S matrices go. A puzzling statement in 
\cite{mp} that one can derive pp wave string  amplitudes from nonextremal 
correlators was analyzed, and we showed that the amplitude is 
actually vanishing in 
the Penrose limit, and to get a nonzero result we are forced to the 
usual (extremal) string field theory calculations 
\cite{cfhmmps,cfhm,kpss,bkpss,sv,svtwo,psvvv},... 

We have defined S matrices on pp waves, generalizing a procedure due to 
Giddings \cite{gid}, first to flat space with nonzero 5 dimensional 
(i.e. AdS) mass and then to pp waves. The procedure turns boundary-to-bulk 
propagators into normalizable wavefunctions, so we have checked that the 
AdS wavefunctions have the correct flat space and pp wave limits (and 
also that pp wave wavefunctions turn into flat space wavefunctions).
There was previously no direct test of the procedure that we are aware 
of. 

We have then tested the procedure on the correlators that we know 
are not renormalized: general scalar 2- and 3-point functions and 
extremal correlators. We have written down the general 3-point function,
but we found that 
it is not obvious how to proceed in the general case (it is quite 
complicated, and the order of limits is highly nontrivial). We have then 
concentrated on the extremal 3-point function, the simplest case we can 
analyze, and also of relevance, since (with appropriate limits on the 
representation of operators) this correlator will survive the Penrose limit.

Taking the flat space limit on it though, we have found a discrepancy. 
The AdS side result is just a delta function, that we have expressed 
in spherical harmonics in order to compare with the SYM result. 
We have found that the angular dependence works, but we get extra 
gamma functions containing numerical $\Delta$ factors, as well as extra 
energy factors, (in d=4) of $(p_2/p_1)^{\Delta_2 -3/2} (p_3/p_1)^{\Delta_3
-3/2}$. We have seen that a similar thing will happen for the general 
extremal correlator (if we apply the flat space limit on it, 
not the pp wave limit!). 

So what could be the reason for the discrepancy? 
Of course, one answer would be to say 
that the procedure is not good. Basically, we are turning bulk to 
boundary propagators into normalizable wavefunctions. But in AdS correlators
the bulk to boundary propagators are integrated over the bulk points, and 
the flat space limit supposes that we sit at finite 5-th coordinate r, while
taking $R\rightarrow \infty$. But since we integrate over all r's, we have 
to make sure that only the contribution of finite r survives. While the 
contribution near the boundary is negligible for normalizable wavefunctions,
there is still a contribution at $r\sim R$ that is still far away from the 
boundary, and it is not obvious that is also small. One could maybe try 
to see how this affects the correlator. Another potential 
problem was already pointed out in \cite{gid}, namely that loop corrections 
will modify the poles $\omega_{nl}$ that we have factorized near (in LSZ 
fashion) by (\ref{giddi}). But we have chosen especially the extremal 
3-point function for the certainty that the free result is exact, so 
at most it could be a question of defining properly the limit (maybe there 
is some subtlety that was missed before). 

A final possibility would be that the flat space limit does not make sense 
on its own, but that the pp wave limit does (and that one needs to go 
first in the pp wave limit, and then maybe to flat space). We have not 
analyzed the pp wave limit on the extremal 
3-point function ($\Delta\sim R^2, n\sim 1, l \sim 1$), and that could still 
give the right result. For the pp wave case, we know that at least the 
first quantized string picture (and the string field theory calculations)
work, so maybe S matrices are also OK. For those calculations, we have 
also seen that we can understand the flat space limit better as a further 
limit of the pp wave (see e.g., \cite{bmn,bn}), so maybe the same applies 
here. In any case, it is clear that further work is needed to define 
S matrices correctly.

{\bf Acknowledgements} We would like to thank Juan Maldacena for pointing
out to us that there is a problem with the definition of massless S 
matrices on the pp wave, and also to Radu Roiban for discussions.
This research was  supported in part by DOE
grant DE-FE0291ER40688-Task A.

\newpage

\renewcommand{\theequation}{A.\arabic{equation}}
\setcounter{equation}{0}

{\Large\bf{Appendix A. AdS holography review}}

\vspace{1cm}

In this Appendix we will review  AdS-CFT correlators and holography
in various coordinates. 
All the original correlator calculations 
 were done in Euclidean space in Poincare coordinates for AdS.
Things are somewhat simpler there, so we will start weith it.

\vspace{.5cm}

{\em Euclidean Poincare AdS}

In Poincare coordinates, 
\be
ds^2=\frac{R^2}{(x_0)^2}((d\vec{x})^2 +(dx_0)^2)
\ee
the bulk to bulk propagator is (from \cite{lt})
\be
G(x,y)= (x_0y_0)^{d/2} \int \frac{d^d k}{(2\pi)^d} e^{i\vec{k}\cdot (\vec{x} 
-\vec{y})} I_{\nu} (k x_0^<) K_{\nu}(k x_0^>)
\ee
and since $I_{\nu}(x)\sim x^{\nu}, K_{\nu}(x)\sim x^{-\nu}$ when $x 
\rightarrow 0$ we have that 
\be
G(x,y) \sim (x_0)^{d/2+\nu}= (x_0)^{2 h_+}=(x_0)^{\Delta}
\ee
and consequently the bulk to boundary propagator is defined as 
\be 
K_{B\partial}
(\vec{x}, x_0; \vec{y})= \lim_{y_0\rightarrow 0} (y_0)^{-\Delta} G(x,y)
\ee
and is then given by (with normalization and notation from \cite{mp})
\be
K_{B\partial}
(\vec{x}, x_0; \vec{y})\equiv <0|\phi^I (\vec{x}, x_0)O^I(\vec{y}|0>
=[A(\Delta_I)]^{1/2} [\frac{x_0}{x_0^2 +(\vec{x}-\vec{y})^2}]^{\Delta_I}
\ee
which near the boundary behaves as $(x_0)^{d-\Delta} \delta(\vec{x}-\vec{y})$
where 
\be
(x_0)^{d-\Delta}= (x_0)^{d/2- \nu}= (x_0)^{2h_-}
\ee
as could have been inferred already from the form of G(x,y). Here 
\be
2h_{\pm}=\frac{d\pm \sqrt{d^2+4m^2R^2}}{2}
\ee
are the solutions of $m^2R^2=2h(2h-d)$ and represent the possible behaviours 
at infinity of the solutions of the plane wave equation $(\Box -m^2)\phi=0$.

Let us now use an abuse of notation and split $\vec{x}$ into $(t, \vec{x})$.
On the boundary, the euclidean $(\Box- m^2)\phi=0$ would imply $k_0^2 +\vec{k}
^2 +m^2=0$, which has no solution for real $k_0$. Since we have one extra 
(nontrivial) dimension, we can have solutions with real $k_0$ and $\vec{k}$.

The regular solution is 
\be
\phi \propto e^{ik_0 t +i\vec{k}\vec{x}} (x_0)^{d/2} K_{\nu}(|k|x_0)\phi_0 
(k_0, \vec{k})
\ee
and as we saw above behaves as the boundary as $x_0^{2h_-}(1+...)+x_0^{2h_+}
(1+...)$, hence the non-normalizable behaviour $x_0^{2h_-}$ dominates. There 
is also a solution with $K_{\nu}$ replaced by $I_{\nu}$, which behaves as 
$x_0^{2h_+}$, so normalizable, but is badly behaved in the bulk (blows up
exponentially at $x_0\rightarrow \infty$). 

\vspace{.5cm}

{\em Holography for Euclidean Poincare AdS}

So there are two solutions of the AdS wave equation, 
a non-normalizable one with $K_{\nu}$ and a normalizable one (at the boundary)
with $I_{\nu}$, but which blows up in the bulk, thus is not good, and we can 
say there is a unique relevant solution. 

We will see shortly that in Lorentzian AdS there are both normalizable and 
nonnormalizable modes, and they are dual to operator VEVs and sources, in 
a precise sense to be defined. But now, there are only non-normalizable 
modes, and as we saw, the bulk to boundary propagator behaved near the 
boundary like such a mode
(or more precisely like a linear combination of these regular modes), 
so these modes generate sources $\phi_0(\vec{y})$ on the boundary via

\be
\phi (\vec{x}, x_0)= \int d^dy K(\vec{x}, x_0; \vec{y}) \phi_0(\vec{y})
= c\int d^d y \frac{x_0^{2h_+}}{(x_0^2 +(\vec{x}-\vec{y})^2)^{2h_+}}
\phi_0(\vec{y})
\sim (x_0)^{d-\Delta}\phi_0(\vec{x})
\ee
Thus boundary correlators in AdS (from $\phi_0(\vec{y})$ derivatives of the 
AdS partition function) are related to correlators on the boundary (from 
$\phi_0(\vec{y})$ derivatives of the SYM partition function).

However,
because of the AdS-CFT dictionary relating euclidean generating functionals
we get also a one-point function (operator VEV) induced by the same source 
$\phi_0$ (so this is not an independent quantity).
\be
<{\cal O}(\vec{x})>_{\phi_0}=-c(2h_+)\int d^d y \frac{\phi_0(\vec{y})}{
|\vec{x}-\vec{y}|^{2(2h_+)}}
\ee

\vspace{.5cm}

{\em Lorentzian Poincare AdS}

As we mentioned, in Lorentzian signature
the situation is a bit more involved due to the presence of normalizable
modes. It was analyzed first in \cite{bkl}, where the general idea was 
put forward, and then in \cite{bklt} where the details were worked out. 

In Lorentzian signature then for $k^2=-\omega^2 +\vec{k}^2>0$ the solution is 
the same as for the Euclidean case. But on the boundary we want $(\Box- m^2)
\phi=0$ which means $-\omega^2 +\vec{k}^2 +m^2=0$, meaning that the relevant 
solution is for $k^2=-\omega^2+\vec{k}^2<0$, 
so for the physical case we need to analytically continue the Euclidean 
solution. There are then two solutions,
\be
\Phi^{\pm}\propto e^{-i\omega t +i\vec{k} \vec{x}}(x_0)^{d/2}J_{\pm \nu}
(|k| x_0)
\ee
when $\nu= \sqrt{d^2 +4 m^2 R^2}/2$ is not integral and $J_{\pm \nu}$ replaced
by $Y_{\nu}$ when it's integral. Since $J_{\nu}(x)\sim x^{\nu}$ around x=0, 
$\Phi^-$ behaves like $(x_0)^{2h_-}$ and is non-normalizable, as the euclidean 
solution. But now we also have $\Phi^+$ which behaves like $(x_0)^{2h_+}$ and 
is normalizable (and also well defined in the interior, unlike in the 
Euclidean case).

The propagators are the obvious analytical continuation of the Euclidean 
Poincare propagators.

\vspace{.5cm}

{\em Holography for Lorentzian Poincare AdS}

Now the dictionary is a bit more involved. As in Euclidean space, the 
non-normalizable modes in the bulk define sources on  the boundary, but now 
one can add normalizable modes in the bulk field which don't affect the source
(have subleading behaviour) but affect the one-point function (operator VEV)
\be
\phi(x_0,\vec{x})=\phi_n(x_0, \vec{x})+
c\int d^d y \frac{x_0^{2h_+}}{(x_0^2 +(\vec{x}-\vec{y})^2)^{2h_+}}
\phi_0(\vec{y})
\ee
where $\phi_n(x_0, \vec{x}) \rightarrow (x_0)^{2h_+}\tilde{\phi}_n(\vec{x})$
and 
\be
<\tilde{\phi}_n|{\cal O}(\vec{x})|\tilde{\phi}_n>_{\phi_0}
= (2h_+)\tilde{\phi_n}(\vec{x})+c(2h_+)\int d^d y \frac{\phi_0(\vec{y})}{
|\vec{x}-\vec{y}|^{2(2h_+)}}
\ee

At the operatorial level, the bulk field (normalizable mode)
\be
\hat\phi_n=\sum_k[a_k\phi_{n,k}+a^{+}\phi^*_{n,k}]
\ee
is thus mapped to the operator (acting as a field)
\be
\hat {\cal O}=\sum_k[b_k \tilde{\phi}_{n,k}+b^+_k\tilde{\phi}_{n,k}^*]
\ee
where we can identify the operators $a_k=b_k$, and the operator acts on 
a coherent state  $|\tilde{\phi}_n>=
e^{ct. b_k^+}|0>$. 

The short form of this statement (from \cite{ps}) is that 
non-normalizable modes are mapped to sources and normalizable modes to VEVs
(or states), so that
\be
\phi\sim a_i (x_0)^{d-\Delta}+b_i (x_0)^{\Delta}
\ee
implies 
\be
H=H_{CFT}+a_i {\cal O}_i
\ee
and 
\be
<0|{\cal O}|0> =b_i {\rm or \;\; rather}\;\; <b_i|{\cal O}_i |b_i>=b_i+
(a_i \; {\rm piece})
\ee

The very important consequence is that if we put the non-normalizable mode 
to zero (no sources) and look at a bulk configuration (probe) which 
corresponds to a combination of normalizable modes (maybe with non-normalizable
components as well), it will get mapped to a VEV of the dual operator. 

In particular, examples of probes studied in \cite{bklt} are D-instantons, 
fundamental and D-strings, and dilaton wavepackets. For instance, the 
D-instanton goes near the boundary as 
\bea
&&e^{\phi}\sim g_s +c \frac{x_0^4 \tilde{x}_0^4 }{[\tilde{x}_0^2 +|\vec{x}
-\vec{y}|^2]^4}...\nonumber\\&&
\chi =\chi_{\infty}\pm (e^{-\phi}-1/g_s)
\eea
and implies a VEV for the operator coupling to $\phi$
\be
\frac{1}{4g_{YM}^2}<Tr F^2(\vec{x})> =\frac{48}{g_{YM}^2}\frac{\tilde{x}_0^4}
{[\tilde{x}_0^2+|\vec{x}-\vec{y}|^2]^4}
\ee
which is just the formula for the YM instanton! Moreover, this is an example 
of scale-radius duality, since the radial position of the D-instanton, 
$\tilde{x}_0$ is mapped to the scale of the instanton.

If we put the normalizable mode to zero instead, we get the same $\phi(\vec{x}
, x_0)$ and $<{\cal O}>|_{\phi_0}$ as in Euclidean Poincare AdS, and 
boundary AdS correlators are related in the same way to SYM correlators.

\vspace{.5cm}

{\em Global Lorentzian AdS}

In global coordinates,
\be
ds^2=R^2(-cosh ^2 \mu dt^2 +d\mu^2 +sinh ^2 \mu d\Omega^2_{d-1})
\ee
or, with $tan \rho= sinh \mu $,
\be
ds^2=\frac{R^2}{cos^2\rho}(-dt^2+d\rho^2+\sin^2 \rho d\Omega_{d-1}^2)
\ee
The boundary is now at $\rho=\pi/2$. The solutions to the wave equation are 
written as
\be
\Phi=e^{-i\omega t}Y_{l, \{ m\}}(\Omega)\chi (\rho) 
\ee
where $Y_l$ are spherical harmonics on $S^{d-1}$, $\nabla^2 _{S^{d-1}}
Y_l=-l(l+d-2)Y_l$, and 
can be analyzed near the origin and the boundary. At the 
boundary, we find solutions $\Phi^{\pm}$ that behave as 
$\Phi^{\pm}\sim (cos\rho)^{2h_{\pm}}$. But at 
the origin, only one of the solutions $\Psi_{1,2}$ that we find
is regular, namely one that can be written as $\Psi_1=C^+\Phi^++C^-\Phi^-$. 

So the unique regular solution is in general non-normalizable (since 
$\Phi^-$ is nonnormalizable), but we get a quantization condition from 
$C^-=0$ for which we get normalizable solutions. The condition is
\be
\omega_{nl}=2h_++2n+l
\ee
where n is a positive integer (or zero). Then the solutions are normalizable
and their asymptotic behaviour is (with the notation in \cite{gid})
\be
\chi_{nl}(\rho) \sqrt{2\omega_{nl}}\rightarrow k_{nl}(cos \rho )^{2h_+}
\ee

So whereas in Poincare coordinates the Lorentzian wave equation has continous
normalizable solutions (as well as non-normalizable ones), in global 
coordinates the general case is non-normalizable, and particular cases are 
normalizable. 

Somewhat similar to the Poincare case, in the global case the bulk to bulk 
propagator is 
\be
iG(x,y)=\int \frac{d\omega}{2\pi} \sum_{n l\vec{m}}
e^{i\omega (t-t')}\frac{\phi^*_{nl\vec{m}}(\vec{x})\phi_{nl\vec{m}}(\vec{y})
}{\omega_{nl}^2-\omega^2 -i\epsilon}
\ee
which again behaves like a normalizable mode towards the boundary, so 
the bulk to boundary propagator is 
\be
K_{B\partial}
(\vec{y},x) =2\nu R^{d-1} lim_{\rho '\rightarrow \pi/2}(cos \rho ')^{2h_+}
iG(x,y)
\ee
and near the boundary behaves like a non-normalizable mode,
\be
K_{B\partial}
(\vec{y}, x)\sim (cos \rho) ^{2h_-}\delta (\vec{x}-\vec{y})
\ee
and can be written as (note that on the boundary -and not only- we parametrize
$\vec{x}=(t, \hat{e})$ where $\hat{e}$ takes values in $\Omega_{d-1}$)
\be
K_{B\partial}(\vec{y},x)=2\nu R^{d-1}
\int \frac{d\omega}{2\pi} \sum_{n l\vec{m}}
e^{i\omega (t-t')}\frac{k_{nl} Y^*_{l\vec{m}}(\hat{e})\phi_{nl\vec{m}}(\vec{x})
}{\omega_{nl}^2-\omega^2 -i\epsilon}
\ee
The propagator can also be rewritten as 
\be
K_{B\partial}
(\vec{y}, x)= K_B[\frac{\cos^2\rho}{[cos^2 (t-t') -sin \rho \hat{e}\hat{e}']^2 
+i\epsilon}]^{h_+}
\ee

\vspace{.5cm}

{\em Holography in global Lorentzian AdS}

Lorentzian global AdS has as a boundary the cylinder $S_3\times R$, and 
the CFT in $R^4$ is mapped to the cylinder, and dimensionally reduced 
on $S^3 $ to a QM Hamiltonian. 

Non-normalizable modes (general frequencies $\omega$)
correspond, as before, to sources 
for the CFT operators, whereas at special frequencies, normalizable modes 
correspond to the states of the CFT on the cylinder.
\be
\omega_{nl}=\Delta + 2n +l
\ee

\vspace{.5cm}

{\em Cylinder $(t, \vec{u})$ AdS}

Finally, \cite{mp} uses a third coordinate system which is important since 
one can take the pp wave limit more easily in it. We will therefore 
write down the sphere part of the gravity metric explicitly as well. 
They make the coordinate 
change from Euclidean Poincare AdS 
\be
e^{\tau}=(x_0^2 + \vec{x}^2)^{1/2}, \;\;\; \vec{u}=\vec{x}/x_0 = u \hat{e}
\ee
where now the boundary of AdS is at $u=\infty$ and is parametrized by
\be
e^{\tau}=|x|, \;\;\; \hat{e}= \vec{x}/ |x|
\ee
followed by the Wick rotation to Lorentzian signature $\tau \rightarrow 
(1-i\epsilon) it$.
One gets the $AdS\times S$ metric (with the sphere written in Wick-rotated 
analogous coordinates)
\be
ds^2 =R^2(-(1+u^2)dt^2 +d\vec{ u}d\vec{u} -\frac{u^2 du^2}{1+u^2})
+R^2 ((1-v^2)d\psi^2+d\vec{v}d\vec{v} +\frac{v^2dv^2}{1-v^2})
\ee
As we mentioned, on the boundary we change coordinates from the plane 
(corresponding to Euclidean Poincare in AdS) to the $S^3\times R$ cylinder
(corresponding to the ($t,\vec{u}$) cylinder as well as global coordinates
in AdS) via $x_i=e^{\tau} \hat{e}_i, \tau= it$.
Then the change of coordinates $u=sinh \mu =tan \rho$ takes us to the usual 
global coordinates defined before. One can also easily check that the 
Poincare bulk to boundary propagator becomes in these coordinates
(just modifying the position of the 
$i\epsilon$) as the one in \cite{mp} (with a different normalization)
\be
K(t, \vec{u}; t', \hat{e}') = \frac{(A(\Delta))^{1/2}}{2^{\Delta}
(\sqrt{1+u^2} cos [(1-i\epsilon)(t-t')]-\vec{u} \hat{e})^{\Delta}}
\ee
Then, in between this coordinate system and the Euclidean 
Poincare coordinate system the only difference is that we need to make a 
conformal transformation on the operator on the boundary, which changes 
from the plane (for Poincare) to the cylinder (for $(t, \vec{u})$ coordinates),
so multiply by $e^{\Delta \tau}=e^{i\Delta t}$. 

A very important observation is that if we put $t'=-\infty$ (or $\vec{y}=0$) 
in the bulk to boundary  propagator 
before making the coordinate change to the $(t, \vec{u})$ cylinder
(actually before the Wick rotation, really), the propagator becomes 
\be
K(t, \vec{u}; -\infty, \hat{e}')= (A(\Delta))^{1/2} [ e^{-it} (1+u^2)^{-1/2}
]^{\Delta}
\label{propwick}
\ee
which is different from what we will get if we just put $t'=-\infty$ 
in the bulk-to-boundary propagator. This will be very important later on
(in the main text).

\renewcommand{\theequation}{B.\arabic{equation}}
\setcounter{equation}{0}

\vspace{1cm}

{\Large\bf{Appendix B. Delta function in spherical harmonics}}

\vspace{1cm}

In this Appendix we rewrite the delta function in a way that could match 
the SYM calculation, namely expanding in spherical harmonics.

The spatial momentum delta function in global AdS parametrization is 
 (the time integral would give the energy conservation, so 
we are left with a 4d space integral in global AdS)
\bea
&&(2\pi)^d
\delta^d(p_1\hat{e}_1+p_2\hat{e}_2-p_3\hat{e}_3)= \int d^d x
e^{i(p_1r\hat{e}_1\hat{e}'+p_2r\hat{e}_2\hat{e}'-p_3r\hat{e}_3\hat{e}')}
\nonumber\\&&
=\sum_{l_1\vec{m}_1l_2\vec{m}_2l_3\vec{m}_3}i^{l_1+l_2+l_3}(2\pi)^{-3d/2}
Y_{l_1\vec{m}_1}(\hat{e}_1)Y_{l_2\vec{m}_2}(\hat{e}_2)Y_{l_3\vec{m}_3}
(-\hat{e}_3)\nonumber\\&&
\int d\vec{e}'Y_{l_1\vec{m}_1}(\hat{e}')Y_{l_2\vec{m}_2}
(\hat{e}')Y_{l_3\vec{m}_3}(\hat{e}')
\nonumber\\
&&\int r^{d-1} dr \frac{J_{l_1+d/2-1}(p_1r)J_{l_2+d/2-1}(p_2r)
J_{l_3+d/2-1}(p_3r)}{(p_1p_2p_3 r^3)^{d/2 -1}}
\eea
where we have actually expressed the exponentials in the form we got them 
from the AdS propagators, as Bessel functions and spherical harmonics 
(the solution in spherical coordinates). Substituting d=4 and generalizing
to arbitrary number of external momenta we get 
\bea
&&\sum_{l_i\vec{m}_i}i^{l_1+l_2+...+l_n}(2\pi)^{-2n}
Y_{l_1\vec{m}_1}(\hat{e}_1)Y_{l_2\vec{m}_2}(\hat{e}_2)...Y_{l_n\vec{m}_n}
(-\hat{e}_n)\int d\vec{e}'Y_{l_1\vec{m}_1}(\hat{e}')Y_{l_2\vec{m}_2}
(\hat{e}')...Y_{l_n\vec{m}_n}(\hat{e}')
\nonumber\\
&&\frac{1}{p_1...p_n}
\int_0^{\infty} r^{3-n} dr J_{l_1+1}(p_1r)J_{l_2+1}(p_2r)...
J_{l_n+1}(p_nr)
\eea
At this point we can use the formula (from \cite{gr})
\bea
&& \int _0^{\infty} 
\prod_j [J_{\mu_j}(b_j x)]\{\cos [(\rho+\sum_j\mu_j -\nu)\pi
/2]J_{\nu} (ax)\nonumber\\&&
+\sin  [(\rho+\sum_j\mu_j -\nu)\pi /2] Y_{\nu}(ax) \}\frac{x^{\rho -1}
}{x^2+k^2}dx =-\prod_j I_{\mu_j}(b_jk)K_{\nu}(ak)k^{\rho-2}
\eea
which applies if $Re(k)>0, a>\sum_j |Re b_j|, Re(\rho +\sum_j \mu_j)>|Re 
(\nu)|$. As we can see we apply this formula for $k\rightarrow 0$ (real)
$b_i=p_i$, $\mu_i=l_i+1$, $\nu=l_n+1$, $a=p_n$, $\rho=6-n$. The first condition
is satisfied, the second is satisfied only as a limit, since $p_n=p_1+...
+p_{n-1}$ is just energy conservation. The last condition becomes 
$4+\sum_{i=1}^{n-1} l_i-l_n >0$, and finally one needs 
$4+\sum_{i=1}^{n-1} l_i-l_n =2m$ (even number) so that we only have the 
$J_{l_n+1}$ term, and not the Y term. Then (using that at $x\sim 0$ 
we have $I_{\nu}(x)\simeq (x/2)^{\nu}1/\Gamma(\nu +1), K_{\nu}(x)
\simeq (x/2)^{-\nu}\Gamma(\nu)/2$) 
\bea
&&\int_0^{\infty} r^{3-n} dr J_{l_1+1}(p_1r)J_{l_2+1}(p_2r)...
J_{l_n+1}(p_nr)\nonumber\\&&
= -\frac{l_n!}{(\prod_{i=1}^{n-1}(l_i+1)!)2^{\sum_i l_i-l_n 
+n-1}}\frac{\prod_j p_j^{l_j+1}}{p_n^{l_n+1}}
\lim _{k\rightarrow 0}k^{\sum_i l_i -l_n+2}
\eea

In the n=3 case, we get 
\be
-\frac{l_3!}{(l_1+1)!(l_2+1)!2^{l_1+l_2-l_3+2}}p_1^{l_1+1}p_2^{l_2+1}
p_3^{-l_3-1}\lim _{k\rightarrow 0} k^{l_1+l_2-l_3+2}
\ee
This formula can be checked against another formula for 3 Bessel 
integral from \cite{gr}, valid for $p_3>p_1+p_2$, so 
still used as a limit in our case $p_3=p_1+p_2$, 
\bea
&&\int_0^{\infty} dx J_{l_1+1}(p_1x)J_{l_2+1}(p_2x)J_{l_3+1}(p_3 x) 
=\frac{p_1^{l_1+1}p_2^{l_2+1}p_3^{-l_1-l_2-3}\Gamma(\frac{l_1+l_2+l_3}{2}+2)}
{(l_1+1)!(l_2+1)!\Gamma (\frac{l_3-l_1-l_2}{2})}
\nonumber\\&&
F_4 (\frac{l_1+l_2-l_3}{2}+1, \frac{l_1+l_2+l_3}{2}+2, l_1+2, l_2+2; 
p_1^2/p_3^2, p_2^2/p_3^2)
\nonumber\\&&
=\frac{p_1^{l_1+1}p_2^{l_2+1}p_3^{-l_1-l_2-3}\Gamma(\frac{l_1+l_2+l_3}{2}+2)}
{(l_1+1)!(l_2+1)!\Gamma (\frac{l_3-l_1-l_2}{2})}
{}_2F_1(\frac{l_1+l_2-l_3}{2}+1, \frac{l_1+l_2+l_3}{2}+2, l_1+2, x)
\nonumber\\&&
{}_2F_1(\frac{l_1+l_2-l_3}{2}+1, \frac{l_1+l_2+l_3}{2}+2, l_2+2, y)
\eea
where $x(1-y)=p_1^2/p_3^2, y(1-x)=p_2^2/p_3^2$, and therefore if $p_3=
p_1+p_2$ we have $x=p_1/p_3$ and $y=p_2/p_3$. One then uses that 
${}_2F_1(0,...)=1$ and ${}_2F_1(-m,...)=$ polynomial 
to check against the general n formula. 

Finally, for the spherical harmonics integrals appearing on both sides 
one should use some technology derived strictly speaking for $S_2$, but 
which should hold in general. The Gaunt formula,
\be
Y_{l_1m_1}(\hat{e})Y_{l_2m_2}(\hat{e})= \sum _l \sqrt{\frac{(2l_1+1)(2l_2+1)}
{4\pi (2l+1)}}
<l_1l_2m_1m_2|lm> <l_1l_2 00|l0>Y_{lm}(\hat{e})
\ee
is derived from the identification of 
\be
Y_{lm}\sqrt{\frac{2l+1}{4\pi}}\bar{D}_{m0}^l(\hat{e})
\ee
where $D_{mm'}^j(R)$ is the representation R of the rotation group acting on 
$\psi_{jm}$. Then one deduces ($Y_{l-m}(\hat{e})=(-)^mY_{lm}^*(\hat{e})$)
\bea
&&\int d\hat{e} Y_{l_1m_1}(\hat{e}) Y_{l_2m_2}(\hat{e})Y_{l_2m_3}(\hat{e})
=(-)^{m_3}\sqrt{\frac{(2l_1+1)(2l_2+1)}{4\pi(2l_3+1)}}
\nonumber\\&&<l_1l_2m_1m_2|l_3m_3>
<l_1l_2 00 |l_3 0>
\eea
or using the Wigner 3-j symbols 
\be
(j_1, j_2, j_3; m_1, m_2, -m_3)= 
(-)^{j_1-j_2+m}/ \sqrt{2j+1}<j_1 j_2 m_1 m_2 | jm>
\ee
it is 
\be
\frac{1}{\sqrt{4\pi}}\sqrt{(2l_1+1)(2l_2+1)(2l_3+1)}
(j_1, j_2, j_3; m_1, m_2, m_3)(j_1, j_2, j_3; 0, 0, 0)
\ee

By repeated application of the Gaunt formula we get 
\bea
&& I_{\{l_i m_i \} }\equiv \int d\hat{e}
Y_{l_1m_1}(\hat{e})Y_{l_2m_2}(\hat{e})...Y_{l_{n-1}m_{n-1}}(\hat{e})
Y_{l_nm_n}(\hat{e})= \frac{1}{(4\pi)^{n/2-1}}\sqrt{(2l_1+1)...(2l_n+1)}
\nonumber\\&&
(2l_2'+1)...(2l_{n-2}'+1)
(l_1 l_2 l_2';m_1, m_2, m_2') (l_1 l_2 l_2' ; 0 0 0) 
\nonumber\\&&(l_2' l_3 l_3' ; m_2' m_3
m_3')(l_2' l_3 l_3'; 0 0 0)...(l_{n-2}' l_{n-1} l_n; m_{n-2}' m_{n-1}
m_n)(l_{n-2}' l_{n-1} l_n; 0 0 0)
\eea
and then 
\bea
&&(2\pi)^4\delta^4 (p_1\hat{e}_1+p_2\hat{e}_2+...-p_n\hat{e}_n)
\nonumber\\&&=
-\sum_{l_i\vec{m}_i}i^{l_1+l_2+...+l_n}(2\pi)^{-2n}
Y_{l_1\vec{m}_1}(\hat{e}_1)Y_{l_2\vec{m}_2}(\hat{e}_2)...Y_{l_n\vec{m}_n}
(-\hat{e}_n)\nonumber\\&&I_{\{l_i m_i \} }
\frac{l_n!}{(\prod_{i=1}^{n-1}(l_i+1)!)2^{\sum_i l_i-l_n 
+n-1}}\frac{\prod_j p_j^{l_j}}{p_n^{l_n+2}}
\lim _{k\rightarrow 0}k^{\sum_i l_i -l_n+2}
\nonumber\\&&
=-\sum_{l_i\vec{m}_i}i^{l_1+l_2+...+l_n}(2\pi)^{-2n}
Y_{l_1\vec{m}_1}(\hat{e}_1)Y_{l_2\vec{m}_2}(\hat{e}_2)...Y_{l_n\vec{m}_n}
(-\hat{e}_n)\nonumber\\&&
I_{\{l_i m_i \} }
\frac{(\sum l_i +2)!}{(\prod_{i=1}^{n-1}(l_i+1)!)2^{n-3}}
\frac{1}{p_n^4}\prod_j (\frac{p_j}{p_n})^{l_j}
\label{deltan}
\eea
and that is the decomposition of the delta function in spherical harmonics, 
which should be matched with SYM. 

In the last line we have replaced $l_n=\sum l_i+2$ to make the power of k =0,
but notice that then strictly speaking the result is zero, since the quantity 
$I_{\{ l_i, m_i \} }$ composes angular momenta, so we see that it satisfies 
the generalized 'triangle' inequalities, in particular $l_n\leq \sum l_i$. 
Moreover, we see that for $l_n > \sum l_i +2$ we have a divergent result 
(even though the summation makes it zero). 

But in order to calculate Bessel function integrals we have to 
take various limits (like the $k\rightarrow 0$ limit), and then 
maybe we have to take into account the effect of high l's.  Indeed, the 
delta function should be zero most of the time. Since we are working for 
$p_n=\sum p_i$, this representation of the delta function should be nonzero 
only if $\hat{e}_i=\hat{e}_n$ for all i. 

But there seems to be only one way that this is achieved: for different 
$\hat{e}_i$'s, we can always find a sufficiently high but finite l above 
which the sum over spherical harmonics averages out (since the 
numerical coefficient will be approximately constant and the spherical 
harmonics themselves will oscillate drastically, like $\sin (l\; \phi)$, for 
instance). Since for finite l's we saw that the sum is zero, the whole result 
is zero at different $\hat{e}_i$'s. But when all $\hat{e}_i$'s are 
exactly the same, we can't find any finite l above which the spherical 
harmonics average out. 

So we will only calculate the contribution from l going to infinity, and 
moreover since now there is no formula for the Bessel function integrals, 
we will assume that the finite l formula holds, except that now we neglect 
the 2 and have $l_n=\sum l_i$, and also keep a possible divergent factor, 
that is, (with $l_i=a_i m, m\rightarrow \infty$)
\bea
&&\frac{1}{p_1...p_n}
\int_0^{\infty} r^{3-n} dr J_{l_1+1}(p_1r)J_{l_2+1}(p_2r)...
J_{l_n+1}(p_nr)\nonumber\\&&
= -\frac{(\sum l_i+2)!}{(\prod_{i=1}^{n-1}(l_i+1)!)2^{n-3}}
\frac{1}{p_n^4}\prod_j (\frac{p_j}{p_n})^{l_j} (1/k)
\eea

Note that \cite{gr} have also the explicit formulas 
\be
\int _0^{\infty} dx x^{\nu-M+1}J_{\nu}(bx)\prod_{i=1}^kJ_{\mu_i}(a_ix)
=0, \;\; M=\sum_i \mu_i
\ee
which applies if $b=p_n>\sum p_i=\sum_i a_i$ (and therefore as a limit for the 
equality case), to give 
\be 
\int _0^{\infty} dr r^{3-n} J_{l_1+1}(p_1r)...J_{l_n+1}(p_nr)=0
\;\;{\rm if }\;\; l_n=\sum l_i
\ee
(if $l_n- \sum l_i\neq 0 $ one adds it to the power of r)
and similarly 
\be
\int _0^{\infty} dx x^{\nu-M-1}J_{\nu}(bx)\prod_{i=1}^kJ_{\mu_i}(a_ix)
=2^{\nu-M-1}b^{-\nu} \Gamma(\nu)\prod_{i=1}^k\frac{a_i^{\mu_i}}{\Gamma
(1+\mu_i)}
\ee
implying
\be
\int _0^{\infty} dr r^{3-n(+l_n-\sum l_i -2)} J_{l_1+1}(p_1r)...J_{l_n+1}(p_nr)
= 2^{l_n-\sum l_i -n+1}\frac{l_n!}{p_n^{l_n+1}}\prod_i \frac{p_i^{l_i+1}}
{(l_i+1)!}
\ee
which is exactly what we obtained from the other formula, for 
$l_n=\sum_i l_i+2$. 

Moreover, we can check directly what happens at large l, using the formula
for large $\nu$
\be
J_{\nu}(\frac{\nu}{\cosh \alpha})= \frac{e^{\nu\tanh \alpha -\nu\alpha}}{
\sqrt{2\pi \nu \tanh \alpha}}\{ 1+\frac{1}{\nu}(1/8\coth \alpha -5/24 \coth^3
\alpha)+...\}
\ee
Use $\nu=x e^{\alpha}$ for $\alpha \rightarrow \infty$, x fixed to get 
\be
J_{\nu}(x/2)\simeq x^{\nu}\frac{e^{\nu}}{\nu^{\nu}\sqrt{2\nu \pi}}
\ee
and therefore at large l's 
\be
\int _0^{\infty}dr r^{3-n}\prod_i J_{l_i+1}(p_i r)= 
\prod_{i=1}^n \left[ \frac{e^{l_i+1}p_i^{l_i+1}}{(l_i+1)^{l_i+1}
\sqrt{2\pi (l_i+1)}
} \right]\int _0^{\infty} dr r^{3-n+\sum (l_i+1)}
\ee
which we will see has a form similar to our extrapolation of the 
finite l Bessel integral formula, up to the divergent factor. We use 
the Sterling formula to find that the product is $\prod_i p_i^{l_i+1}/(l_i+1)!
$ (including n). Putting the exponent of r to -1 (least divergent), thus 
formally $l_n+1=-(\sum_i l_i+3)$, gives the same formula as from the finite l 
extrapolation, with an indeterminate factor $(\int dr/r)/\Gamma (0)$.

Going back to our extrapolated formula,
the numerical coefficient of the $l\vec{m}$ sum at large $l$ 
can be evaluated using the Sterling formula
\be
l_j!\simeq (l_j+1)^{l_j+1/2}e^{-l_j-1}\sqrt{2\pi}
\ee
Then (substituting $l_i=ma_i$)
\bea
&&\frac{1}{p_n^42^{n-3}}[\frac{l_n!}{\prod_i (l_i+1)!}\prod_j (\frac{p_j}{
p_n})^{l_j}]=\nonumber\\&&
\frac{(2\pi)^{1-n/2}}{p_n^42^{n-3}}e^{
(ma_n+1/2)\ln (ma_n+1)-\sum_i (ma_i+3/2)\ln (ma_i+2)-l_n+\sum l_i +2n-3
+\sum_j l_j ln (p_j/p_n)}
\eea
and the exponent becomes, in an m expansion (introducing also a nonzero 
$b_i= l_i-ma_i$ and $c=l_n -ma_n$ for use further on, but skipped in the 
following equations)
\bea
&&m \ln  m (a_n-\sum a_i)+m(a_n \ln a_n -\sum_i a_i \ln a_i 
+\sum _j a_j \ln p_j/p_n+\sum_i a_i-a_n)\nonumber\\&&
-\frac{ \ln m}{2} (3n-4+2\sum_i b_i -2c)+\frac{1}{2}(\ln a_n - 3
\ln (\prod_i a_i))+ c ln \; a_n \nonumber\\&&
-\sum_i b_i ln \;a_i+\sum_j b_j ln 
\frac{p_j}{p_n}
\eea
If we substitute $a_n=\sum a_i$ (since we need $l_n\leq \sum l_i$, but 
for finite l that gives zero, so we will define $a_n=\sum a_i$, and 
the l's can differ by a finite amount), the exponent is 
\be
m f(a_i)-\frac{\ln m}{2} (3n-4)+\frac{1}{2}(\ln (\sum a_i)-3\ln (\prod a_i))
\ee
where 
\be
f(a_i)=(\sum_i a_i)\ln (\sum_i a_i)-\sum_i a_i \ln (a_i)+\sum_j a_j 
\ln (p_j/p_n)
\ee
then the maximum of f is defined by 
\be
\frac{\partial f}{\partial a_i}=\ln (\frac{a_n p_i}{a_i p_n})=0
\rightarrow \frac{a_i}{a_n}=\frac{p_i}{p_n}\Rightarrow f(a_{i, max})=0
\ee
since then $\partial ^2f/\partial a_i^2 =1/a_n-1/a_i<0$. 

Since we are dealing with the coefficient of a divergent quantity (m), it 
makes sense to keep only the values in the sum over l's where the coefficient 
is maximum, so we substitute $a_i/a_n=p_i/p_n$. But then the m term vanishes, 
and we are left with only ln m and constant terms in the exponent. But now 
note that these terms depend on how we define $a_i$'s, in particular whether 
$l_n=\sum l_i +c$ with nonzero c (or $l_n=m(\sum a_i)+c$), and also
whether $l_i=ma_i+b_i$. With nonzero c and $b_i$ the exponent is  
\be
-\ln m \frac{3n-4-2c+2b_i}{2}-\frac{1}{2}
\ln \frac{\prod_i (a_i)^{3+2b_i}}{a_n^{1+2c}}+\sum_jb_j \ln \frac{p_j}{p_n}
\ee
and so 
\bea
&&(2\pi)^4 \delta^4 (p_1\hat{e}_1+p_2\hat{e}_2+...-p_n\hat{e}_n)
\nonumber\\&&=
-\sum_{l_i\vec{m}_i}i^{l_1+l_2+...+l_n}(2\pi)^{-2n}
Y_{l_1\vec{m}_1}(\hat{e}_1)Y_{l_2\vec{m}_2}(\hat{e}_2)...Y_{l_n\vec{m}_n}
(-\hat{e}_n)\nonumber\\&&I_{\{l_i m_i \} }
\frac{(2\pi)^{1-n/2}}{p_n^4 2^{n-3}}[m^{c+2-3n/2-\sum_ib_i}\frac{a_n^{c+1/2}}
{\prod_i (a_i)^{3/2+b_i}}]\prod_j (\frac{p_j}{p_n})^{b_j}
\label{deltasph}
\eea
Notice that if one takes $c=3n/2 -2+\sum_ib_i$ 
two nice things happen: the divergent m 
factor dissapears and the square brackets become only momentum dependent
(using $a_i/a_n=p_i/p_n$)
\be
[\prod_i(\frac{p_n}{p_i})^{3/2}]
\ee

\renewcommand{\theequation}{C.\arabic{equation}}
\setcounter{equation}{0}

\vspace{1cm}

{\Large\bf{Appendix C. Identities for limits of wavefunctions}}

\vspace{1cm}

For the flat space and pp wave limits of wavefunctions in 
section 4, we need to find the limits for 
\be
\lim_{n\rightarrow \infty}
\frac{1}{n^{\alpha}} P_n^{\alpha, \beta} (cos (x/n)) {\rm \;\; and}\;\;
\lim_{n\rightarrow \infty}
\frac{1}{n^{\alpha}} P_n^{\alpha, n \beta} (cos (x/n))
\ee
where $P_n^{\alpha, \beta}$ are Jacobi polynomials, since these appear in 
the AdS wavefunctions, as well as the limits of spherical harmonics of large 
l for the sphere wave functions, as well as the limits for Hermite polynomials
for the case of flat space limit of the pp wave.

\be
P_n^{\nu,\mu}(x)=\frac{(-)^n}{2^n n!}(1-x)^{-\nu}(1+x)^{-\mu}\frac{d^n}{dx^n}
\{ (1-x)^{\nu +n}(1+x)^{\mu +n}\}
\label{jacobi}
\ee
Then for $x\rightarrow 1-x$ and expanding in powers in the brackets and 
taking derivatives we get 
\be
P_n^{\nu, \mu}(1-x)=
\frac{(1-x/2)^{-\mu}}{n!}\sum_{m\ge 1}\frac{(-x/2)^m}{m!}(\mu +n)...
(\mu +n -m +1)(\nu+n+m)...(\nu+m+1)
\ee
Then one gets 
\be
\lim_{n\rightarrow \infty}
\frac{1}{n^{\alpha}} P_n^{\alpha, \beta} (cos (x/n))=
\sum_{m\ge 1}\frac{(-)^m(x/2)^{2m}}{m! \Gamma(\alpha + m+1)}
=(\frac{2}{x})^{\alpha}J_{\alpha}(x)
\ee
and similarly 
\be
\lim_{n\rightarrow \infty}
\frac{1}{n^{\alpha}} P_n^{\alpha, n \beta} (cos (x/n))
= \sum_{m\ge 1}\frac{(-)^m(\sqrt{1+\beta}\frac{x}{2})^{2m}}{m! \Gamma(\alpha
 + m+1)}=(\frac{2}{x\sqrt{1+\beta}})^{\alpha}J_{\alpha}(x\sqrt{1+\beta})
\ee

For the Hermite polynomials one has
\bea
&& \lim_{n\rightarrow \infty}\frac{1}{n^{\alpha}}L_n^{\alpha}(x/n)
=x^{-\alpha /2}J_{\alpha}(2\sqrt{x})\nonumber\\
&& H_{2n}(x)=(-)^n 2^{2n}n! L_n^{-1/2}(x^2)\nonumber\\
&& H_{2n+1}(x)=(-)^n 2^{2n+1}n! x L_n^{1/2}(x^2)
\eea
which implies
\bea
&&\lim_{n\rightarrow \infty}\frac{(-)^n\sqrt{n}}{2^{2n}n!}H_{2n}(\frac{x}
{\sqrt{n}})=\sqrt{\frac{2}{\pi}}cos (2x)\nonumber\\
&&\lim_{n\rightarrow \infty}\frac{(-)^n}{\sqrt{n}2^{2n+1}n!}H_{2n+1}(\frac{x}
{\sqrt{n}})=\sqrt{\frac{2}{\pi}} sin (2x)
\eea

For the pp wave limit of AdS wavefunctions, we first write an alternative 
version for $P^{\nu, \mu}_n(1-x)$. We replace $x\rightarrow 1-x$ in 
(\ref{jacobi}) and first take derivatives (and not expand), we get 
\be
P_n ^{\nu, \mu}(1-x)=\frac{1}{2^n} \sum_{m=0}^n \begin{pmatrix} n+\nu&\\
m&\end{pmatrix}\begin{pmatrix} n +\mu &\\n-m&\end{pmatrix} (-x)^{n-m} (2-x)^m
\ee
Then one gets 
\be
\lim_{a\rightarrow \infty} P_n^{\nu, \alpha a}(\cos \frac{2r}{\sqrt{a}})
= \sum _{m=0}^n \begin{pmatrix} n+\nu &\\n-m&\end{pmatrix}
\frac{(-\alpha r^2)^m}{m!}= L_n^{\nu} (\alpha r^2)
\label{pplimit}
\ee
where $L_n^{\nu} (x)$ are Laguerre polynomials. 

The d-dimensional harmonic oscillator in spherical coordinates has the 
Schrodinger equation
\be
(-\Delta + \mu^2 \omega^2 r^2 -2\mu E)\Phi= (-\frac{\partial^2}{\partial r^2}
+\frac{d-1}{r}\frac{\partial}{\partial r} +\frac{1}{r^2}\Delta_{S_{d-1}})\Phi
=0
\ee
with the separated solution
\be
\Phi_{nl \vec{m}}= \chi_{nl}(r)Y_{l\vec{m}}(\hat{e})
\ee
and the radial equation
\be
\{ -[\frac{d^2}{dr^2}+ \frac{d-1}{r}\frac{d}{dr}-\frac{l(l+d-2)}{r^2}]
+\mu^2\omega^2 r^2-2\mu E_n \} \chi_{nl}(r)=0
\ee
Separating the small r behaviour $\chi \sim r^{\alpha}\Rightarrow \alpha = 
l$ and the large r behaviour $\chi = e^{-r^2\mu \omega /2}$, i.e. 
\be
\chi = r^l e^{-r^2\mu \omega /2} G(r)
\ee
one finds for G in $z=\sqrt{r}$ variables the equation 
\be
z G'' +G'[ l+ d/2- z (\mu \omega)] + g [ \frac{\mu E_n}{2}- \frac{\mu \omega 
}{2}(l+ d/2)]=0
\ee
which is the Laguerre equation, provided we have the quantization condition
\be
E_n - \omega (l+d/2)= 2n
\ee
thus the d dimensional harmonic oscillator in spherical coordinates has the 
wavefunctions
\be
\Phi_{nl\vec{m}}= N_{nl}r^l e^{-\mu \omega r^2/2}L_n^{l+d/2-1}(\mu \omega 
r^2)Y_{l\vec{m}}(\hat{e})
\label{wavefct}
\ee

\renewcommand{\theequation}{D.\arabic{equation}}
\setcounter{equation}{0}

\vspace{1cm}

{\Large\bf{Appendix D. General 3-point function of scalars}}

\vspace{1cm}

Take
\be
\int \frac{dt_1dt_2dt_3}{(2\pi )^3}e^{i(p_1t_1+p_2t_2+p_3t_3)}
\frac{a_{123}e^{i(k_1t_1+k_2t_2+k_3t_3)}}{|x_{12}|^{2\alpha_3}|x_{13}|^{2
\alpha_2}|x_{23}|^{2\alpha_1}}
\ee
Using that $k_1t_1+k_2t_2+k_3t_3= \alpha_1(t_2+t_3)+\alpha_2(t_1+t_3)
+\alpha_3(t_1+t_2)$ and then using (\ref{equ}) and (\ref{bdprop})
as well as shifting as usual by $t_1$ and making the $t_1$ integral to get 
the delta function we get 
\bea
&& a_{123}\delta(p_1+p_2+p_3)\int \frac{d\omega_1}{2\pi}\frac{d\omega_2}{2\pi}
\frac{d\omega_3}{2\pi}
\int \frac{dt_{21}}{2\pi} e^{it_{21}(p_2-\omega_3+\omega_1)}
\int \frac{dt_{31}}{2\pi} e^{it_{31}(p_3-\omega_2-\omega_1)}\nonumber\\&&
\sum_{n_3l_3\vec{m}_3}
\frac{k_{n_3l_3}^2Y_{l_3\vec{m}_3}^*(\hat{e}_1)Y_{l_3\vec{m}_3}(\hat{e}_2)}
{\omega_3^2- \omega_{n_3l_3}^2-i\epsilon}
\sum_{n_1l_1\vec{m}_1}
\frac{k_{n_1l_1}^2Y_{l_1\vec{m}_1}^*(\hat{e}_2)Y_{l_1\vec{m}_1}(\hat{e}_3)}
{\omega_1^2- \omega_{n_1l_1}^2-i\epsilon}
\sum_{n_2l_2\vec{m}_2}
\frac{k_{n_2l_2}^2Y_{l_2\vec{m}_2}^*(\hat{e}_1)Y_{l_2\vec{m}_2}(\hat{e}_3)}
{\omega_2^2- \omega_{n_2l_2}^2-i\epsilon}
\nonumber\\&&
=a_{123} \delta(p_1+p_2+p_3)\int \frac{d\omega_1}{2\pi}
\sum_{n_3l_3\vec{m}_3}
\frac{k_{n_3l_3}^2Y_{l_3\vec{m}_3}^*(\hat{e}_1)Y_{l_3\vec{m}_3}(\hat{e}_2)}
{(p_2+\omega_1)^2- \omega_{n_3l_3}^2-i\epsilon}
\nonumber\\&&
\sum_{n_1l_1\vec{m}_1}
\frac{k_{n_1l_1}^2Y_{l_1\vec{m}_1}^*(\hat{e}_2)Y_{l_1\vec{m}_1}(\hat{e}_3)}
{\omega_1^2- \omega_{n_1l_1}^2-i\epsilon}
\sum_{n_2l_2\vec{m}_2}
\frac{k_{n_2l_2}^2Y_{l_2\vec{m}_2}^*(\hat{e}_1)Y_{l_2\vec{m}_2}(\hat{e}_3)}
{(p_3-\omega_1)^2- \omega_{n_2l_2}^2-i\epsilon}
\nonumber\\&&
=a_{123} \delta(p_1+p_2+p_3)
\sum_{n_1 l_1 n_2 l_2 n_3 l_3}k_{n_1l_1}^2k_{n_2l_2}^2 k_{n_3l_3}^2
\nonumber\\&&
(\sum_{\vec{m_1}}Y^*_{l_1\vec{m}_1}(\hat{e}_2)Y_{l_1\vec{m}_1}(\hat{e}_3))
(\sum_{\vec{m_2}}Y^*_{l_2\vec{m}_2}(\hat{e}_1)Y_{l_2\vec{m}_2}(\hat{e}_3))
(\sum_{\vec{m_3}}Y^*_{l_3\vec{m}_3}(\hat{e}_1)Y_{l_3\vec{m}_3}(\hat{e}_2))
\nonumber\\&&
[\frac{1}{2\omega_{n_3l_3}}\frac{1}{(p_2 \mp \omega_{n_3l_3})^2-
\omega_{n_1l_1}^2}
\frac{1}{(p_2+p_3 \mp \omega_{n_3l_3})^2-\omega_{n_2l_2}^2}\nonumber\\&&
+\frac{1}{2\omega_{n_1l_1}}\frac{1}{(p_2 \pm \omega_{n_1l_1})^2-
\omega_{n_3l_3}^2}
\frac{1}{(p_3 \mp \omega_{n_1l_1})^2-\omega_{n_2l_2}^2}\nonumber\\&&
+\frac{1}{2\omega_{n_2l_2}}\frac{1}{(p_3+p_2 \mp \omega_{n_2l_2})^2
-\omega_{n_3l_3}^2}
\frac{1}{(p_3 \mp \omega_{n_2l_2})^2-\omega_{n_1l_1}^2}]
\label{three}
\eea
where in the second line we did the $t_{21}, t_{31}, \omega_2, \omega_3$
integrals and in the last line also the $\omega_1$ integral
and the $\pm, \mp$ indicate a sum of terms over the two values. Here 
\be
k_{nl}^2=\frac{2\omega_{nl}}{R^{d-1}}\frac{\Gamma(n+\alpha+1-\frac{d}{2})
\Gamma(n+l+\alpha)}{n!\Gamma(n+l+\frac{d}{2})}\frac{1}{\Gamma(\alpha +1 -\frac{
d}{2})^2}
\label{knl}
\ee
and at large R in the \cite{gid} case (large n, fixed $\alpha$ and l) we have 
\be
\frac{k_{nl}^2R^{d-1}}{2\omega_{nl}}\sim \frac{n^{2\alpha -d}}{\Gamma (\alpha +
1-\frac{d}{2})^2}=(\frac{ER}{2})^{2\nu}\frac{1}{\Gamma(\nu +1)^2}
\ee
Note that doing the sums over m's we could rewrite the spherical harmonics 
as $Y_{l_1}(\hat{e}_2\cdot \hat{e}_3)Y_{l_2}(\hat{e}_1\cdot \hat{e}_3)
Y_{l_3}(\hat{e}_1\cdot \hat{e}_2)$, where $Y_l(\cos \alpha)$ is $P_l 
(\cos \alpha)$.

\end{document}